\documentclass[fleqn,usenatbib]{mnras}

\usepackage{newtxtext,newtxmath}
\usepackage{xcolor}  

\usepackage[T1]{fontenc}

\DeclareRobustCommand{\VAN}[3]{#2}
\let\VANthebibliography\thebibliography
\def\thebibliography{\DeclareRobustCommand{\VAN}[3]{##3}\VANthebibliography}
\newcommand{\HI} {H\,{\sc i}} 

\usepackage{graphicx}	
\usepackage{amsmath}	
\usepackage{booktabs}   
\usepackage{comment}
\usepackage{subcaption}
\usepackage{hyperref}
\usepackage{tabularx}

\title[Short title, max. 45 characters]{MNRAS \LaTeXe\ template -- title goes here}

\title [Baryon census]{The baryon census and the mass-density of stars, neutral gas, and hot gas as a function of halo mass}

\author[Dev et al.]
{Ajay Dev,$^{1,2}$\thanks{E-mail: ajay.dev@research.uwa.edu.au}
Simon P. Driver,$^{1}$
Martin Meyer,$^{1,2}$
Aaron Robotham,$^{1,2}$
Danail Obreschkow,$^{1,2}$ \and
Paola Popesso,$^{3}$ 
Johan Comparat$^{4}$
\\\\
{}$^1$International Centre for Radio Astronomy Research, University of Western Australia, M468, 35 Stirling Highway, Perth, WA 6009, Australia.\\
{}$^2$ARC Centre of Excellence for All Sky Astrophysics in 3 Dimensions (ASTRO 3D), Australia.\\
{}$^3$European Southern Observatory, Karl Schwarzschildstrasse 2, 85748, Garching bei München, Germany\\
{}$^4$Max-Planck-Institut f\"{u}r extraterrestrische Physik (MPE), Giessenbachstrasse 1, D-85748 Garching bei M\"unchen, Germany\\
}

\date{Accepted XXX. Received YYY; in original form ZZZ}

\pubyear{2015}

\begin{document}
\label{firstpage}
\pagerange{\pageref{firstpage}--\pageref{lastpage}}
\maketitle

\begin{abstract}
We study the stellar, neutral gas content within halos over a halo mass range $10^{10} \text{ to } 10^{15.5} \text{M}_\odot$ and hot X-ray gas content over a halo mass range $10^{12.8} \text{ to } 10^{15.5} \text{M}_\odot$ in the local universe. We combine various empirical datasets of stellar, \HI\ and X-ray observations of galaxies, groups and clusters to establish fundamental baryonic mass vs halo mass scaling relations. These scaling relations are combined with halo mass function to obtain the baryon densities of stars, neutral gas and hot gas ($T>10^6 \text{K}$), as a function of halo mass. We calculate the contributions of the individual baryonic components to the cosmic baryon fraction. Cosmic stellar mass density ($\Omega_\text{star}=2.09^{+0.21}_{-0.18} \times 10^{-3}$), cosmic \HI\ mass density ($\Omega_\text{\HI}=0.49^{+0.25}_{-0.12} \times 10^{-3}$) and cosmic neutral gas mass density ($\Omega_\text{neutral gas}=0.71^{+0.39}_{-0.18} \times 10^{-3}$) estimates are consistent with previous more direct method measurements of these values, thereby establishing the veracity of our method. We also give an estimate of the cosmic hot plasma density ($\Omega_\text{hot gas}=2.58^{+2.1}_{-0.66} \times 10^{-3}$). 
\end{abstract}

\begin{keywords}
galaxies: haloes -- galaxies: groups: general -- galaxies: stellar content -- X-rays: galaxies: clusters
\end{keywords}



\section{Introduction}

Dark energy, dark matter and baryonic matter make up nearly all ($>99.99$\%) of the matter-energy density of the Universe at our current epoch \citep{Planck2018}. While these three primary components have been robustly quantified in terms of their contribution to the matter-energy density, we only have a theoretical physics-based understanding of the baryonic components (plasma, gas, stars etc) . 

At the present time, the baryon density constraints inferred from Big Bang Nucleosynthesis (BBN) and Planck Cosmic Microwave Background (CMB) measurements find $\Omega_\text{b}\text{h}^2= 0.022$ \citep{Planck2018}. This global constraint needs to be reconciled with our local baryon census in which we sum the directly detected baryonic components, i.e., stars, cold gas, hot gas, dust etc. 

Historically, most attempts to build a baryon census start from a galaxy-by-galaxy basis, i.e., one first accounts for the baryons bound to galaxies, and later infers the unbound baryons which are generally thought to reside in a more diffuse phase. 

\citet{Fukugita&Peebles2004} presented an extensive inventory of the cosmic energy density associated with nearly 40 distinct components compiled from several observational studies and theoretical estimates. The major baryon components were found to be: stars ($\Omega_\text{stars}=0.0026 \pm 0.0009$), neutral gas ($\Omega_\text{\HI}=(4.2 \pm 0.7)\times 10^{-4}, \Omega_{\text{H}_2}=(1.6 \pm 0.6)\times 10^{-4}, \Omega_{\text{He}}=(2.2 \pm 0.9)\times 10^{-4}$) and intracluster plasma ($\Omega_\text{cl}=0.0018 \pm 0.0007$). They assumed a total baryon density estimate of $\Omega_\text{b}\text{h}^{2} = 0.0225 \pm 0.0015 $ by averaging the baryon density estimates from Wilkinson Microwave Anisotropy Probe (WMAP; \citealt{Bennett2003, Spergel2003}), Sloan Digital Sky Survey (SDSS; \citealt{York2000, Tegmark2004}) and primordial elemental abundance measurements \citep{Kirkman2003, Izotov2004}. 

The remaining baryons which remained unaccounted for from their census (i.e., 90\%), were {\it assumed} to reside in a warm intergalactic plasma. Similar analyses have previously been reported by \citet{Persic&Salucci1992} and \citet{Fukugita1998}. 
More recently, \citet{Shull2012} updated the baryon census by including contributions of the warm-hot intergalactic medium (WHIM) and Lyman-$\alpha$ forest, albeit with large uncertainties. The presence of this unaccounted baryons in a diffuse ionised phase has been confirmed through different tracers, albeit each of the techniques probe a specific phase space of the gas, both in emission (e.g. \citealt{Tanimura2019, deGraff2019, Macquart2020}) and absorption (e.g. \citealt{Prochaska2011, Tumlinson2013, Werk2014, Nicastro2018, Mathur2023}).

One possible improvement over the galaxy-based approach, and the basis of this paper, is to explore the baryon budget by starting from a {\it halo-based} approach. The argument is that it is the halo, that may contain one or several galaxies, which is the fundamental building block of our Universe. 

The density of halos in the Universe is a relatively well understood distribution,  typically given by the halo mass function (HMF; \citealt{Press&Schecter1974, Sheth&Tormen2002, Murray2018}) which describes the number density of halos as a function of halo mass. If the HMF is expressed in the form of a mass density rather than number density, it represents the mass distribution as a function of halo mass. \textcolor{black}{ Cosmological N-body simulations are now widely used to study the HMF \citep{Warren2006, Reed2007, Tinker2008, Watson2013, Bhattacharya2011, Ishiyama2015}.} \citet{Driver_hmf_2022} recently produced an empirical measurement of the halo mass function, based on the combination of Galaxy And Mass Assembly Survey (GAMA; \citealp{Driver2011, Driver_GAMA_2022, Liske2015}) , SDSS and ROSAT-ESO Flux Limited X-ray Galaxy Cluster Survey (REFLEXII; \citealt{Bohringer2013}) data, which was empirically constrained to a halo mass of $10^{12.7} \text{ M}_\odot$. 

The HMF (empirical or theoretical) shows that $\sim 40 \%$ of the total matter density of the Universe is located in the halo mass range of $10^{12}-10^{14} \text{ M}_\odot$, which is called the group regime and $\sim 12\%$ in the cluster regime (i.e., $>10^{14} \text{ M}_\odot$) \citep{Driver_hmf_2022}. Hence, one would expect a similar fraction of the baryon budget to be associated with detectable clusters and groups, and within these environments it should be possible to determine the phases in which matter exists as a function of halo mass. However, the actual distribution of baryons would be biased from the dark matter distribution due to various baryonic and other non-gravitational processes \citep{Duffy2010, Chan2015, Schaller2015, Chisari2018}. In this work, we will use a halo-based approach, to first determine scaling relations for the hot plasma, gas and stars versus halo mass and then combine these with the HMF to recover the total baryon density associated with each component.

Some previous studies have also attempted a halo-based approach to look at baryon fractions, although the studies mainly focussed on the hot ionised plasma and stellar fractions of halos with masses greater than $10^{13} \text{ M}_\odot$. For example, \citet{Gonzalez2013} looked at the stellar and hot-gas components of 12 galaxy clusters and groups at $z\sim0.1$ having $M_\text{h}=(1-5) \times10^{14} \text{ M}_\odot$. \citet{Akino2022} studied the stellar and gas fractions of 136 XXL groups in a mass range of $10^{13}-10^{15} \text{ M}_\odot$ at $0 \le z \le 1$. Both studies found that halos below $\sim 10^{14.5} \text{ M}_\odot$, appear to have baryon fractions significantly lower than the cosmic baryon fraction. In this work, we will extend the baryon fraction studies over a larger range of halo mass and study the distributions of the major baryonic components - neutral gas, stars and hot gas. 

In our halo-based approach, we will first review the existing literature to establish the key baryonic scaling relations - stellar, \HI\ and hot gas mass as a function of halo mass. The stellar-to-halo mass relation (SMHM) links the total stellar mass within a halo to the total mass of the dark matter halo \citep{Moster2010, Weschler&Tinker2018}. It is an important tool to understand galaxy evolution as it is a measure of star formation efficiency across different halo masses and environments. It also points to the various feedback mechanisms that act at different halo mass scales. At the low-mass group-regime, previous studies total group stellar content using halo abundance matching or occupation distributions have found that the group stellar fraction peaks around $\text{log}_{10}(M_\text{h}/\text{M}_\odot) \sim 12 $ \citep{Moster2010, Leauthaud2012}.  

The \HI-to-halo mass relation (HIHM) links the total mass of atomic hydrogen within a halo to the total mass of the dark matter halo \citep{Padmanabhan2017}. Neutral atomic hydrogen is the underlying fuel for star formation and hence its distribution at different halo scales would provide key insights into galaxy formation. Unlike the SMHM, the HIHM trend seems to be less understood. Theoretical studies of the HIHM have been done using hydrodynamical models such as IllustrisTNG \citep{Villaescusa-Navarro2018} or semi-analytical models like SHARK \citep{Chauhan2021} \textcolor{black}{and \citet{Calette2021}. The trend in the HIHM relation have been quite varying among the different models ranging from a double power-law like trend in \citet{Villaescusa-Navarro2018} to a dip in $10^{12}-10^{13} \text{ M}_\odot$ seen in \citet{Chauhan2021}, and a corresponding flattening seen by \citet{Calette2021} in the same halo mass regime with increasing \HI\ mass trends on either side with halo mass.}

Observationally, stacking techniques have been primarily used to understand the average trend of the HIHM relation \citep{Guo2020, Rhee2023, Dev2023}. \textcolor{black}{ The scatter in the relation has also been recently studied both for the centrals \citep{Dutta2022, Korsaga2023, Saraf2024} and the total \HI\ content \citep{Hutchens2023}.} Unlike the SMHM and HIHM, the X-ray plasma-to-halo mass relation (XPHM) have generally been restricted to halo masses above $ 10^{13} \text{ M}_\odot$ and mainly only in galaxy clusters, as it is difficult to detect X-ray emissions from lower mass halos. Few stacking measurements of X-ray emission from the IGM/CGM of galaxy halos have been done to estimate the hot gas mass in lower mass halos \citep{Li2018, Popesso2023}. Galaxy groups being the host of the largest fraction of matter in the Universe along with the mix of the various non-gravitational processes involved in them should thus be an ideal test bed in improving our understanding of the baryon physics and galaxy formation \citep{Lovisari2021}.

Finally, by combining our scaling relations with the HMF, we can estimate the cosmic baryon density of stars, \HI\ and hot gas. Estimates of the cosmic stellar densities can be done using more precise methods such as galaxy stellar mass function \citep{Rodriguez-Puebla2020, Driver_GAMA_2022} Similarly, \HI\ mass densities in the local universe can be estimated using luminosity density and \HI\ mass-to-light ratio \citep{Zwaan2005, Rodriguez-Puebla2020, Rhee2023}. Hence, the stellar and \HI\ cosmic densities can provide a good check of our method. However, the total cosmic plasma density estimates are quite uncertain as previously these have been studied mainly in the cluster regime with halo masses greater than $10^{14} \text{M}_\odot$. \citet{Fukugita1998} estimated the contribution of intracluster plasma using cluster mass functions and plasma mass fractions in halos above $10^{14} \text{M}_\odot$ with an additional estimation from  groups by extrapolating the cluster mass function using 18 ROSAT X-ray group plasma mass fraction measurements \citep{Mulchaey1996}. In this work, we aim to better constrain the total hot gas density, albeit this would also involve some extrapolation of the XPHM relation to lower halo masses.  

This paper is organised as follows. Section 2 describes the various datasets we use to establish the three key scaling relations (SMHM, HIHM and XPHM). In Section 3, we describe our methodology for estimating the baryonic component densities by combining the scaling relations with the HMF. The results and the various caveats are discussed in Section 4 and in Section 5 we present our conclusions. 

For consistency with previous studies, we consider clusters as structures with halo masses above $10^{14} \text{M}_\odot$, groups to be in the halo mass range $10^{12} - 10^{14} \text{M}_\odot$ and lower mass halos to be predominantly isolated halos with a single galaxy. The halo mass definition used in this paper refers to the total mass of a halo within a spherical overdensity of 200 times the critical density of the Universe $M_{200c}$. However, we note that different measurement techniques use different halo mass criterion, so we try to use the best available halo mass consistent with our definition and apply relevant corrections when needed.
Throughout the paper, we refer to the hot gas/plasma/X-ray component to be ionised gas, with a temperature ($T \geq 10^{6}$K) sufficient to produce X-ray emissions. We assume a Hubble constant of $H_\text{o} = 70 \text{ km s}^{-1} \text{Mpc}^{-1}$ and a flat $\Lambda$CDM cosmology with $\Omega_\text{M} = 0.3, \Omega_\Lambda = 0.7$, and correct for the cosmology in various datasets as necessary.

\section{The baryonic scaling relations}
In this section we look to construct the fundamental scaling relations defined by stellar mass, neutral gas mass and hot gas mass versus halo mass. In the following subsections we consider each of these baryonic components in turn and describe the various datasets used, followed by our fitting of these relations.

\subsection{The stellar-mass halo-mass relation (SMHM)}
To determine the SMHM relation we use a combination of literature datasets as shown on Fig.\,\ref{fig:scaling_rels}(a). For the group regime, we use the data (black open circles) from the GAMA survey. GAMA is a spectroscopic survey conducted on the Anglo Australian Telescope of $\sim300k$ galaxies covering a total of $\sim230 \text{deg}^2$ of sky spread across five regions, and extending
 down to an $r$-band limiting magnitude of $r<19.8$ mag. GAMA data have been made available through a number of Public releases and here we make use of GAMA Data Release 4\footnote{\url{http://www.gama-survey.org/dr4/}} \citep{Driver_GAMA_2022}. We use the GAMA galaxy group catalogue (G$^3$C; {\sc G3CFoFGroupv10}) which was constructed  using a friends-of-friends algorithm applied to the GAMA dataset \citep{Robotham2011}. G$^3$C contains over $26\text{k}$ groups with 2 or more members, and  extends to a redshift of $z \approx 0.4$. \textcolor{black}{From G$^3$C, we select all groups that have 10 or more spectroscopically confirmed group members (i.e., multiplicity or $N_{\rm FoF} \geq 10$).} \if This equates to 3061 groups located in the G09$^h$, G12$^h$, G15$^h$ and G23$^h$ regions. \fi The multiplicity cut of $N_{\rm FoF} \geq 10$ is motivated by the increasing halo-mass uncertainty with decreasing multiplicity (see \citealt{Driver_hmf_2022} figure\,3). At $N_{\rm FoF} = 10$, we estimate the halo mass uncertainty, based on \cite{Driver_hmf_2022}, to be $\Delta \log_{10} M_\text{h} = \pm 0.2 \text{ dex}$ which we adopt as the maximum halo mass error for all selected GAMA groups. We make a further redshift cut of $z<0.1$, to restrict the study to low redshift, resulting in a sample of 102 GAMA groups. The halo masses of the GAMA groups ({\sc MassAfunc}) are derived empirically from the group's velocity dispersion using the {\sc GAPPER} algorithm \citep{Beers1990} as described in \cite{Robotham2011}. These halo masses have been calibrated to simulated mock halo catalogues. The GAMA dynamical masses have also been shown to be consistent with caustic mass estimates within a factor of 2 for more than 90\% of the GAMA groups \citep{Alpaslan2012}.The stellar-masses of the individual group members are calculated based on spectral energy distribution (SED) fitting of the multi-wavelength GAMA photometry (see \citealt{Bellstedt2020}) using the SED fitting code {\sc ProSpect} \citep{Robotham2020, Thorne2021}. 
 
 The {\it initial} stellar-mass for each GAMA group is taken as the direct sum of the individual stellar masses of the confirmed group members. In order to correct for the stellar mass of galaxies that have not been detected due to the magnitude limit of the survey, we make a correction based on the total galaxy stellar mass function (GSMF) reported in \citet{Driver_GAMA_2022}. We adopt the double Schecter fit of the GSMF taken from Table\,7 of \citet{Driver_GAMA_2022}. We rescale the {\it initial} stellar-mass of each group upwards by the ratio of the GSMF integrated down to a cutoff mass and the total integral of the GSMF. The cutoff mass for each group, which depends on its redshift, indicates the distance at which the GAMA data is 99.5 percent complete for $M_*<10^{10} \text{ M}_\odot$. We adopt the cutoff mass from the mass completeness limit curve in Fig.\,10 of \citet{Driver_GAMA_2022}. The typical correction factor applied to most of the groups is $< 10 \%$ (i.e., on the whole this correction is fairly modest and hence the majority of the stellar mass of the groups has been captured from the detected members). 

To further populate the group and cluster regime, we include measurements based on the Sloan Digital Sky Survey \citep{York2000}. For this purpose, we use the group catalog of \citealt{Tempel2014} which is mainly based on SDSS Data Release 10 (DR10; \citealt{Ahn2014}). A FoF algorithm was implemented in \citet{Tempel2014} to identify the galaxy groups. We use the flux-limited group catalogue and the corresponding galaxy catalogue, to identify the group members, from the Catalog Archive Server of SDSS\footnote{\url{https://vizier.cds.unistra.fr/viz-bin/VizieR?-source=J/A+A/566/A1}}. The halo masses in the catalog have been calculated based on the velocity dispersion of the groups. The total stellar mass of each group is estimated by summing the stellar mass of the group members.\textcolor{black}{ The stellar masses, taken from the {\it wisconsin\_pca\_bc03-26.fits}  file\footnote{\url{https://www.sdss4.org/dr17/spectro/galaxy_wisconsin/}}, are calculated from the optical rest-frame spectral region using a principal component analysis method (PCA) as described in \citet{Chen2012}. We use the stellar masses calculated based on the stellar population synthesis model of \citet{Bruzual&Charlot2003} with a \citet{Kroupa2001} initial mass function.}

\textcolor{black}{To be consistent with the stellar mass completeness limit of GAMA survey, we restrict the SDSS groups below $z<0.04$. Similar to the correction applied to GAMA groups, we also correct the stellar masses of SDSS groups to include the contribution of galaxies below the magnitude limit of SDSS survey. We only include groups which have halo masses above $10^{12} \text{M}_\odot$ and have 10 or more group members. This results in 475 galaxy groups/clusters spanning a halo mass range of $10^{12}$ to $ 10^{15} \text{M}_\odot$ (shown by the brown crosses).}

To extend the SMHM relation to lower halo masses we include observational data from \citet{Posti2019} (yellow triangles). This contains 110 disc galaxies taken from the SPARC sample of galaxies \citep{Lelli2016} with high quality \HI/$\text{H}\alpha$ rotation curves and near-infrared Spitzer photometry. The halo masses for these galaxies are calculated by fitting \HI\ rotation curves assuming a NFW spherical halo model for the dark matter distribution. The rotation curve modelling has been done with a Bayesian approach with three free parameters - halo mass, concentration and stellar mass-to-light ratio hence the halo mass is a direct output of the fit. The initial analysis was performed on 158 galaxies, and the final selection excludes those systems which either had poor NFW halo model fits to the observed rotation curve, or had a multi-modal or flat posterior on the halo mass (see \citealt{Posti2019}). Table\,A.1 of \citet{Posti2019} provides the halo mass and the stellar fraction ($f_* = M_*/f_\text{b} M_{\text{halo}}, \text{where} f_\text{b}=0.188$) for the full sample of 110 galaxy groups. The corresponding stellar and halo mass errors are also taken from the table. \textcolor{black}{We also restrict the SPARC sample to include only those systems which have halo masses $< 10^{12} \text{M}_\odot$, resulting in 84 galaxies. In this halo mass regime, the satellite stellar mass fraction is predicted to be less than $10\%$ \citep{Calette2021}. Hence, the inclusion of only central stellar mass through SPARCs sample in the low halo mass regime has negligible impact in our final stellar-halo mass scaling relation. }

Finally, at the higher halo mass end where GAMA and SDSS data contains relatively few high-mass clusters, we include 8 rich clusters from the Sydney-Australian astronomical observatory Multi-object Integral-field spectroscopy (SAMI) cluster redshift survey \citep{Owers2017} (green diamonds). The typical group multiplicity ranges from 50-200 members in these clusters. The halo mass of these clusters were calculated using the velocity dispersion and the quoted halo masses have been scaled up by a factor of $\times 1.25$ as recommended in \citet{Owers2017} to fairly compare with GAMA halo masses (see their Section\,4.2). For the SAMI cluster stellar masses, we sum up the stellar masses of the individual cluster members for each cluster. The stellar mass values are taken from the SAMI input catalogue for the cluster regions ({\sc SAMI InputCatClustersDR3}). The intracluster light (ICL) have been claimed to contribute notably to the total stellar luminosity or stellar mass of clusters ($M_\text{halo}>10^{14} \text{M}_\odot$), although the observed ICL fraction varies anywhere from 0-50\% with large uncertainties \citep{Contini2021, Montes2022}. As we sum up the stellar masses of individual galaxies to obtain the total stellar mass of a halo, we do not include the ICL contribution. Hence, to include the ICL contribution, we increase the calculated stellar masses of our clusters ($M_\text{halo}>10^{14} \text{M}_\odot$) by an additional 20\%. This has a negligible effect on our final results. 

Fig.\ref{fig:scaling_rels}(a) shows the composite stellar-mass halo mass scaling relation for galaxy groups from the datasets described above. In general the datasets appear to be consistent and overlapping albeit with significant scatter, and spans a halo mass range of $10^{10}-10^{15} \text{ M}_\odot$. The halo mass errors are in general larger than the stellar mass errors, with the errors in halo mass most closely related to group multiplicity \citep{Robotham2011}. \textcolor{black}{We note that in these studies, the halo mass measurement technique varies from the low and high halo mass regime, and could introduce a bias. Unfortunately, the use of a heterogeneous sample, at this point in time, is unavoidable. }

\subsection{The neutral-gas mass halo-mass relation (HIHM)}
To study the HIHM relation, we include literature studies which are based on group \HI\ masses calculated by summing individual galaxy \HI\ contributions, or overall group \HI\ measurements. \citet{Guo2020} empirically measured the \HI\ content using a group-spectral stacking technique. They also used \HI\ data from the ALFALFA survey and selected groups from the SDSS-based group catalogue of \citet{Lim2017a}. \citet{Dev2023} performed a similar group-spectral stacking of GAMA groups \citep{Robotham2011} using \HI\ data from the ALFALFA survey, to look at the average group \HI\ content as a function of halo mass. Both the group spectral stacking results are consistent with each other.  

\citet{Rhee2023} used the ASKAP DINGO early-science data \citep{Johnston2007, Johnston2008, Hotan2021, Meyer2009} to estimate the \HI\ content in galaxy groups from the $ G23^h$ region of the GAMA survey. The \HI\ mass was measured by stacking the \HI\ spectra of individual galaxy members in each group selected from within halo mass bins. This measurement will likely be a lower limit of the group \HI\ content as it would miss out on any \HI\ residing outside the galaxies or from \HI\ sources that are optically undetected. \citet{Hutchens2023} used RESOLVE \citep{Kannappan2008} and ECO surveys \citep{Moffett2015} to study the group \HI-halo mass relation by summing up the galaxy \HI\ content of all the group members in their group catalogue. We use their mean HIHM relation for this work (shown in Fig. 17 of \citealt{Hutchens2023}). They report a median scatter of $\sim 0.3$ dex in the relation.  

The combined literature measurements, outlined above, probe the \HI\ content in the halo mass regime of $10^{11}-10^{15} \text{M}_\odot$. To extend to one more dex lower in halo masses, we include data from SPARC \citep{Lelli2016}. These are isolated galaxies for which \HI\ mass are measured directly from their \HI\ emission data, and the halo masses are calculated by fitting their rotation curves with a dark matter density profile \citep{Posti2019}. The same dataset has been described in Section 2.1. \citet{Korsaga2023} used the SPARC sample and observed a universality in the \HI-halo mass ratio for isolated disk galaxies. We also include eleven galaxies from the GMRT-based (Giant Metrewave Radio Telescope) survey called GARCIA (GMRT archive atomic gas survey; \citealt{Biswas2022}). \citet{Biswas2023} performed mass-modelling to derive the halo masses of these eleven galaxies using HI data from GMRT, infrared data from Spitzer and optical data from SDSS. \textcolor{black}{We restrict the SPARC and \citet{Biswas2023} samples to $M_\text{h}<10^{12} \text{M}_\odot$. In this regime, the satellite \HI\ mass fraction is less than $10\%$ \citep{Calette2021, Chauhan2020}. Hence, the inclusion of only central \HI\ mass in the low halo mass regime has negligible impact in our final \HI-halo mass scaling relation. We also bin the SPARC and \citet{Biswas2023} datasets and use the mean \HI\ mass and halo mass in a given bin, for consistency with other datasets while fitting.}

\textcolor{black}{We also show two studies of HIHM relation constructed empirically, although we do not include these in our fitting of the scaling relation.} \citet{Obuljen2019} constructed the \HI\ mass function (\HI MF) by matching detected ALFALFA sources to optically selected galaxy group members from the SDSS group catalogue of \citet{Yang2007}. \citet{Li2022} constructed an \HI\ mass estimator which links several galaxy properties to an \HI-to-stellar mass ratio. This estimator was then used to construct a conditional \HI MF which was integrated to obtain the HIHM relation for groups. The mass function for each group was constructed using a 2D stepwise maximum likelihood estimator with a Schechter function fit. The \HI\ mass was computed by integrating the \HI MF. We also display a number of individual measurements of notable nearby systems including: the MW, M31, Hydra and Virgo, with the \HI\ masses taken from \citet{Kalberla2009}, \citet{Chemin2009}, \citet{Wang2021} and \citet{Li2022}, respectively. The halo masses for these 4 systems have been taken from \citet{Piffl2014}, \citet{Kafle2018}, \citet{Wang2021} and \citet{Kashibadze2020}, respectively. 

\begin{figure*}
  \includegraphics[scale=0.27]{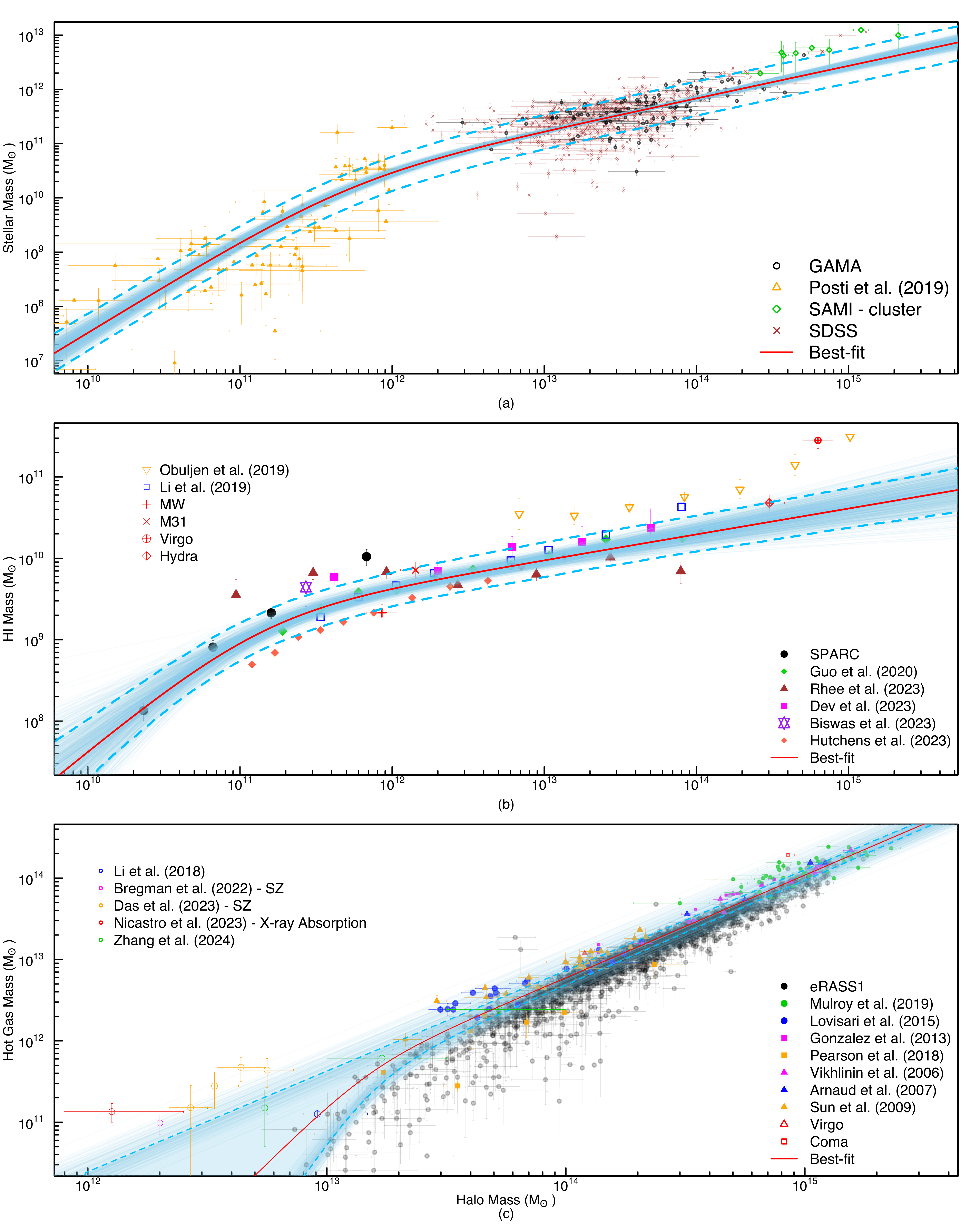}
  
 \caption{Baryonic scaling relations - a) The stellar-mass halo-mass relation from literature data described in Section\,2.1. A double power-law fit has been shown in red with the posterior sampled curves in blue. b) The \HI-mass halo-mass relation from literature data described in Section\,2.2. c) The hot gas-mass halo-mass relation from literature data described in Section\,2.3. Double power-law fits has been shown in red with the posterior sampled curves for the best-fit relation in blue on each panel. The blue dashed curves show 16th and 84th percentile of the total scatter in the relation. For the hot gas-halo mass relation, the dashed blue lines show a single and double power-law fits without and with eRASS1 data respectively. And the best-fit red curve is taken as the mean of the two fits. In all the above sub-panels, only the datasets described in the bottom-right legend are used for the fitting of the relation. }
 \label{fig:scaling_rels}
\end{figure*}

Fig.\,\ref{fig:scaling_rels}(b) shows the \HI\ mass in groups as a function of halo mass for the various \HI\ studies described above. The errors for the datapoints are taken from their respective sources. Overall the combined datasets appear to agree well, and the intrinsic scatter within individual datasets comparable to the scatter between datasets. Towards the higher halo mass end ($> 10^{14} \text{M}_\odot$), we note that the HIHM relation is not well constrained due to the relatively limited number of very nearby rich clusters with robust \HI\ mass measurements. The \HI\ measurements include a combination of group and individual galaxy measurements which was necessary to sample a wide halo mass range. 

\begin{table*}
\centering
\begin{tabular}{cccccc}
\toprule
& \multicolumn{4}{c}{\textbf{Double power-law Fit}} &
                 \\
\midrule
& $\alpha$        & $\beta$       &  $\mathcal{M}_o$        &  $N$   & $\sigma$
               \\ 
                
\midrule
\textbf{SMHM}    &    $0.60 \pm 0.03$	&    
                      $1.71 \pm 0.11$	  &    
                      $11.72 \pm 0.12$   &   
                      $10.47 \pm 0.12$   &  
                      $0.29 \pm 0.01 $ \\
\textbf{HIHM}    &   $0.32 \pm 0.06$	&    
                     $1.54 \pm 0.35$	  &     
                     $11.13 \pm 0.28$   &    
                     $9.38 \pm 0.20$   &    
                     $0.23 \pm 0.03$ \\
\textbf{XPHM}    &   $1.24 \pm 0.01$	&    
                     $3.10 \pm 0.16$	  &     
                     $13.09 \pm 0.03$   &    
                     $11.68 \pm 0.04$   &    
                     $0.072 \pm 0.001 $ \\
\bottomrule
\end{tabular}
\caption{Best-fit double power-law parameters with their errors for the SMHM, HIHM and XPHM relations. $\alpha$ and $\beta$ are the power-law slopes, $\mathcal{M}_o$ is the turn-over halo mass, $N$ is the normalization and $\sigma$ is the predicted total scatter in the relation from the bayesian fitting.}
	\label{tab:best_fits}
\end{table*}

\subsection{The X-ray (hot) plasma mass halo-mass relation (XPHM)}
For the hot gas component of groups, we look to literature measurements from previous group or cluster studies studies at X-ray wavelengths. Here we adopt data from the following studies - \citet{Arnaud2007, Vikhlinin2006, Sun2009, Lovisari2015, Pearson2017, Mulroy2019, Bulbul2024}.

\citet{Arnaud2007} used XMM-Newton data of 10 relaxed clusters in  a redshift range of $z=0.03-0.16$ to study the $M_{500}-Y_X$ relation, where $Y_X$ is the product of the X-ray temperature and gas mass. The halo masses ($M_{500}$) were derived from {\sc nfw} fits to the mass profiles measured down to overdensities of $\Delta_\text{obs}=600-700$ for 8 clusters and $\Delta_\text{obs}\sim 1400$ for 2 clusters, thus there is some extrapolation of the {\sc nfw} model to obtain the $M_{500}$ estimates. The gas masses were derived from the gas density profiles, which were obtained from the emissivity corrected surface brightness profiles using the deprojection and PSF-deconvolution technique of  \citet{Croston2006}.

\citet{Vikhlinin2006} studied the gas and mass profiles of 13 clusters with Chandra observations. Gas and total masses for the clusters are obtained by modelling the observed X-ray surface brightness and projected temperature profiles with analytical functions assuming hydrostatic equilibrium. A modified $\beta$-model was used for the three-dimensional gas density distribution, and a 9-parameter model was used to fit the temperature profiles. An {\sc nfw} model was used for deriving the total cluster mass at $R_{500}$. Table\,4 of \citet{Vikhlinin2006} provided the total mass (${M}_{500}$) and gas fraction (${f}_{\text{gas},500}$) for 10 out of 13 clusters. We multiplied the gas fraction with total mass to obtain the gas mass. 

\citet{Gonzalez2013} used XMM-Newton X-ray data to study the hot gas content for 12 galaxy clusters at $z\sim0.1$ in the halo mass range of $M_{500} = 1-5 \times 10^{14} \text{ M}_\odot$. Halo masses were estimated from M-T relation from \citet{Vikhlinin2009} with the temperatures measured from spectral-fitting. Gas masses were estimated from fitting the surface brightness profile with a $\beta$-model and integrating the resultant gas density profile. We use the values for the gas mass and halo mass reported in Table\,2 of \citet{Gonzalez2013}.

\citet{Sun2009} studied 43 galaxy groups between $z=0.012-0.12$ using archival Chandra data. An 11-parameter gas density model was assumed which provided a good fit for all the groups in their sample. The gas density profile was converted to an emissivity profile using the {\sc mekal} model within {\sc xspec}. The observed surface brightness profile was fitted to the projected emissivity profile along the line of sight to obtain the density parameter values. The group temperature profiles were fit with the same model used in \citet{Vikhlinin2006}. An {\sc nfw} profile was adopted to derive the total mass from the mass density profiles. The gas masses were robustly measured up to ${R}_{500}$ for 11 groups and the results were extrapolated for another 12 groups from properties measured from $\ge {R}_{1000}$. Table\,3 of \citet{Sun2009} provided the total mass (${M}_{500}$) and gas fraction (${f}_{\text{gas},500}$) for 23 out of 43 groups. We multiplied the gas fraction by the total mass to obtain the gas mass.

\citet{Lovisari2015} studied the group X-ray scaling relations using a complete sample of groups from ROSAT based catalogs - NORAS and REFLEX (\citeauthor{Bohringer2000} \citeyear{Bohringer2000} and \citeauthor{Bohringer2004} \citeyear{Bohringer2004} respectively). The X-ray observations for these groups were taken using XMM-Newton and had a redshift range of $z=0.01-0.035$. The surface brightness profiles were fit with single and double $\beta$-models. The extra component model was adopted by looking at the degrees of freedom and $\chi^2$ values. The extracted spectra were fit with an absorbed APEC thermal plasma model \citep{Smith2001}. Temperature profiles were modelled using the parametrizations proposed by \cite{Durret2005} and \cite{Gastaldello2007} and that with the best $\chi^2$ chosen. Gas density profiles were fitted with a double $\beta$-model. The total mass was calculated assuming hydrostatic equillibrium. Table\,2 of \citet{Lovisari2015} provides the group mass (${M}_{500}$) and gas mass (${M}_{\text{gas},500}$) for 20 groups.

\citet{Pearson2017} used Chandra observations to study the gas properties of 10 optically selected galaxy groups from the GAMA galaxy group catalogue. Initial group mass estimates were based on group r-band luminosity which was used to estimate over-density masses ($M_{500}$) using scaling relations calibrated to dark-matter simulations. The APEC thermal plasma model was again used to fit the group region to estimate a source temperature after background removal. The fitted temperature was used to estimate a revised group mass (${M}_{500}$) and radius (${R}_{500}$) using the M-T and r-T relations of \citet{Sun2009}. Single or double $\beta$-model fits to the surface brightness profile were used to estimate the gas number density profile which was then integrated to $R_{500}$ to obtain the total gas mass of the group. Table\,2 of \citet{Pearson2017} provides the group mass (${M}_{500}$) and gas mass fraction (${f}_{\text{gas},500}$) for all 10 groups. We multiply the gas fraction with total mass to obtain the gas mass for each group.

\citet{Mulroy2019} studied a sample of 41 X-ray selected clusters from the 'High-$L_X$' sample of the Local Cluster Substructure Survey which is based on the ROSAT All Sky Survey Catalogues. These selected clusters, spanning a redshift of $0.15<z<0.3$, were studied over a wide range of wavelengths covering X-ray, optical, near-infrared, and millimeter wavelengths with X-ray observations from both Chandra and XMM-Newton. The cluster masses (${M}_{500}$) and radii (${R}_{500}$) were based on weak-lensing masses from \citet{Okabe&Smith2016}, and calculated by fitting an NFW mass profile to the shear profile obtained from Subaru/Suprime-Cam observations. Gas masses were estimated from spherical integrals of the gas density profiles calculated by fitting the X-ray surface brightness profiles. Tables\,1 \&\,2 of \citet{Pearson2017} provides the group mass (${M}_{500, \text{WL}}$) and gas mass (${M}_{\text{gas},500}$) for the 41 clusters.

\citet{Bulbul2024} studied the X-ray properties of 12\,247 optically confirmed galaxy groups and clusters detected in the first western galactic half of the eROSITA All-Sky Survey (eRASS1, \citet{Merloni2024}). Preliminary processing of these data were performed using the eROSITA Science Analysis Software System. 
The initial system radii ($R_{500}$) were estimated based on the L-M scaling relation from \citealt{Chiu2022}, which was later improved using a forward-modelling approach which involved rigorous treatment of the X-ray background and other corrections. X-ray analysis of these systems were performed using the MBProj2D code \citep{Sanders2018} which used images and exposure maps in multiple energy bands to calculate the X-ray properties like temperature and density. Gas masses within $R_{500}$ were measured by integrating the gas electron density assuming spherical symmetry and the density profile model from a modified \citet{Vikhlinin2006} model without the second $\beta$ component. The total mass of the system ($\text{M}_{500}$) was estimated using the scaling relations between count-rate and weak lensing shear measurements described in detail in \citealt{Ghirardini2024}. The groups and clusters in eRASS1 extend to redshifts of $\sim z>1$. As we are focussed on the low-redshift scaling relation, we make a redshift cut of $z < 0.2$, resulting in 3\,413 eRASS1 systems. Additionally, to increase the purity and reduce the contamination of the sample we apply cuts similar to the cuts applied to generate the cosmology sample in eRASS1 (i.e., $\text{L}_\text{ext}$>6 and $\text{P}_\text{cont}$<0.5, c.f. \citealt{Bulbul2024} for details on the definitions), which finally results in 2\,239 eRASS1 systems that we use in this study.

Finally, we include individual cluster measurements for Coma and Virgo. \citet{Simionescu2017} studied the Virgo cluster using Suzaku observations and estimated a virial mass of $M_{500} = (8.3 \pm 0.1) \times 10^{14} \text{ M}_\odot$ and a gas mass of $M_{\text{gas}, 500} = (0.83 \pm 0.01) \times 10^{14} \text{ M}_\odot$. \citet{Mirakhor2020} studied the Coma cluster using XMM-Newton along with Planck Sunyaev Zel'dovich effect observations to recover a halo mass of $M_{200} = (8.50 \pm 0.55) \times 10^{14} \text{ M}_\odot$ and a gas mass of $M_{\text{gas}, 200} = (1.33 \pm 0.09) \times 10^{14} \text{ M}_\odot$.

Fig.\,\ref{fig:scaling_rels}(c) shows the hot gas mass detected in X-ray as a function of halo mass for the X-ray datasets described above and spanning a halo mass range of $10^{12.8}-10^{15.4} \text{ M}_\odot$. The errors for all the datapoints are taken from their respective sources. The halo mass and gas masses are calculated within a virial radius of ${R}_{500}$. In order to make the halo masses consistent with the stellar and \HI\ studies, we convert the halo masses of these groups from ${M}_{500}$ to ${M}_{200}$ using the scaling relation provided in \citet{Ragagnin2021} which were constructed using Magneticum simulations. We use the functional form which converts masses between different overdensities without any assumption of their concentration and profile (c.f. section 5.2 of \citealt{Ragagnin2021} for further details). For all our groups, the mean ratio between the two virial masses is $M_{500}/M_{200} \sim 0.69$. We have used the same relation to scale ${M}_\text{gas,500}$ to ${M}_\text{gas,200}$. 

The eROSITA data seems to be offset marginally low with respect to the other X-ray measurements by $> 0.1 \text{dex}$. This may be because of various factors such as the difference in the energy band of extracted observables and/or calibration differences between various X-ray telescopes \citep{Schellenberger2015}. \citet{Bulbul2024} reported that the eROSITA luminosities are, on average 15\% lower compared to Chandra ones. This would likely translate to some offset in the gas masses as well. Another possibility is that the other smaller datasets (non-eROSITA), which are primarily X-ray selected, could be biased towards brightest X-ray sources at a given halo mass, whereas the larger eROSITA sample would be more complete and include lower luminosity sources as well. There is also a potential bias arising from the halo mass measurement techniques in the various datasets (scaling relation calibrated halo masses in eROSITA versus hydrostatic/velocity dispersion based measurement in others). A combination of all these reasons could in principle explain the offset, however, in-depth calibration studies combined with an accurate selection function are needed to properly estimate a correction factor (if any) to be applied in our scaling relation. For the moment we elect to adopt the data as cited in the literature, rather than make any ad hoc corrections. We note that this will ultimately inflate our final errors in a manner that reflects this uncertainty at this time.

The XPHM relation is deficient of data below a halo mass of $< 10^{13} \text{M}_\odot$. Although we do not have group X-ray measurements below this halo mass, a few hot gas measurements from the circum-galactic medium (CGM) of galaxies have been performed in this halo mass regime. \citet{Sanskriti2023} used the thermal Sunyaev-Zel'dovich (tSZ) effect measurements in CGM of over 600k galaxies by cross-correlating the galaxy catalog of {\sc WISE} and {\sc SuperCosmos} with Compton-y maps from Atacama Cosmology Telescope and Planck. They report stacked hot gas measurements in a halo mass range of $10^{11.8} - 10^{12.8} \text{ M}_\odot$ with a mean redshift range of $z=0.2-0.3$. We show 4 of their measurements that are constrained at more than 90\% confidence in Fig.\, \ref{fig:scaling_rels}. These are taken from Table 1 (rows 4-7 with stellar mass binsize of 0.3 dex) of \citet{Sanskriti2023}. \citet{Li2018} studied the baryon content in the CGM by stacking X-ray measurements of 6 local ($d < 100 \text{ Mpc}$) isolated massive ($M_* \ge 1.5 \times 10^{11} \text{ M}_\odot$) spiral galaxies. They construct a fiducial galaxy that has the average properties of the stacked galaxies and then provide an estimate of the baryon budget for the fiducial galaxy. It has a halo mass of $\text{log}_{10 } (M_{200}/\text{M}_\odot) = 12.96 \pm 0.21$, and the hot CGM mass accounting for ($7.8 \pm 3.6$)\% of the cosmic baryon fraction \citep{Komatsu2009}. \citet{Nicastro2023} report the detection of OVII absorption in the stacked X-ray spectra of three L* galaxies (stellar mass $M_* = 10^{10.53} \text{ M}_\odot$, virial temperature $T_\text{vir} \sim 10^6 \text{ K}$, halo mass $M_\text{h}=10^{12.1} \text{ M}_\odot$) with a CGM hot gas mass estimate of $M_\text{hot-CGM}=(1-1.7) \times 10^{11} \text{ M}_\odot$. \citet{Bregman2022} performed SZ stacking of 12 L* spiral galaxies to detect the hot CGM gas content. They report a hot gas ($T_\text{avg} \sim 3 \times 10^6 \text{ K}$) mass of $M_\text{hot-CGM}=0.98 \pm 0.28 \times 10^{11} \text{ M}_\odot$ for an average halo mass of $M_\text{h}=2 \times 10^{12} \text{ M}_\odot$. \citet{Zhang2024} studied the extended-CGM emission out to $\text{R}_\text{vir}$ using eROSITA by stacking halos around galaxies selected from the SDSS-based galaxy and group catalogue, and correcting for the contamination from satellites, AGN and X-ray binaries. It is important to note that the eROSITA stacked measurements at the lower mass end have higher gas mass than the individual eRASS1 detections. However at the low halo mass end, we would expect eROSITA direct detections to be the most gas-rich halos. This discrepancy could be due to systematics in the measurement of gas/halo masses for the detections or stacks. We do not include the CGM points in our fitting of the XPHM relation as some of the given CGM mass estimates could include contributions from multi-temperature gas phases. Since we have sufficient detections in X-ray, we also do not include any stacked results to avoid any potential systematic bias that would arise by combining detections and stacks together. 

\subsection{Analytic fits to the SMHM, HIHM and XPHM scaling relations}
We now derive the baryonic scaling relation for each of the baryonic mass components described in the above sections as a function of halo mass, and using the literature data described above and shown in Fig.\,\ref{fig:scaling_rels}, i.e. SMHM, HIHM and XPHM relations. 

We test a number of methods for fitting the data, including: smooth-splines, polynomial, single and double power-laws. While each of the methods give a reasonable fit to the data we ultimately decided to adopt the double power-law form due to its easy parameterisation. We note similar $\chi^2$ were obtained from all fitting methods and hence not consider the adoption of a double power-law as in any way limiting. We use the following form of the double power-law function to fit to our data,

\begin{equation}
    \mathrm{log_{10}}(M) = N - \mathrm{log_{10}} \left[ \left( \frac{10^{\mathcal{M}_h}}{10^{\mathcal{M}_o}}\right)^{-\alpha} + \left( \frac{10^{\mathcal{M}_h}}{10^{\mathcal{M}_o}}\right)^{-\beta}\right] ,
\end{equation}

where $\mathcal{M}_h = \mathrm{log_{10}}(M_h/\text{M}_\odot)$, $M$ is the baryonic component mass, $\alpha \text{ and } \beta$ are the two exponents of the power-law, $\mathcal{M}_o$ is the turnover halo mass which separates the two different power-law regimes, N is the normalization parameter and it is equal to $\text{log}_{10}(2M)$ at the turnover mass ($\mathcal{M}_h = \mathcal{M}_o$). 

We fit the double power-law using the {\sc brms} package in R which provides a flexible interface for fitting bayesian regression models using \textsc{Stan} \citep{brms, rstan}. Normal priors were specified on the fit parameters and the prior information were based on fits done using least-squares on the same dataset. 
All the fits are inverse-variance weighted based on the errors of the individual data points. 

We show the datasets and the scaling relation fits in Fig.\,\ref{fig:scaling_rels}. The red line indicates the best-fit double power-law and the blue curves show the posterior samples. 

Fig.\,\ref{fig:scaling_rels}(a) shows the SMHM relation with a double power-law fit.\if which has been physically motivated (see \citealt{Behroozi2019}) as the separation of the two halo mass regimes where different feedback mechanisms dominate - AGN at the high halo mass regime and supernovae at the low halo mass regimes.\fi This turnover halo mass also roughly separates groups from single occupancy halos. 

Fig.\,\ref{fig:scaling_rels}(b) shows the HIHM relation which also has been fit with a double power-law similar to the SMHM relation. The turnover halo mass in for HIHM is slightly below the SMHM relation and a little sharper. 

Fig.\,\ref{fig:scaling_rels}(c) shows the XPHM. Before the release of the eRASS1 dataset, majority of the previous hot gas-halo mass relations were primarily X-ray selected and existed in the high mass group and cluster regimes only \citep{Eckert2021, Oppenheimer2021}. The data were typically fit with a single power-law. However, with the larger sample provided by eROSITA, we can observe a deviation from a single power-law as we move from the cluster to group regime. This turn-down appears to be inconsistent with the lower halo mass CGM data.  The CGM measurements seem consistent with a single-power law extrapolation of the XPHM. 

To accommodate this inconsistency, we perform two fits. We fit a double power-law for the XPHM relation for the entire sample {\it excluding} the CGM data. We also fit the older data combined without the CGM and eRASS1 data with a single power-law. These two fits and their errorbands are shown as the dotted blue lines and shading in Fig.\,\ref{fig:scaling_rels}(c). 
Hence for the XPHM relation we take the conservative approach of adopting the mean of the two fits (shown as red line in Fig.\,\ref{fig:scaling_rels}(c)). To bracket the range of uncertainty, we adopt the single and double power-laws given by the dashed blue curves as our upper and lower bounds.

We note that in the near future eRASS4 data will provide a much larger sample in the group regime which would help us to confirm and constrain the turnover mass better. It is also worth noting that X-ray emission from the majority of the halos in the group regime are arguably undetected even by eROSITA \citep{Popesso2023}. Hence, the mean hot gas mass in this regime may in reality be lower. 
However, the non-eRASS1 dataset alone strictly favours a single power-law which, on extrapolation, also agrees with the CGM scale measurements of the hot gas in the low halo mass range (shown by open circles in Fig.\,\ref{fig:scaling_rels}(c)). In a future paper (Dev in prep.), we will perform X-ray stacking of optically detected halos to better constrain this relation in the group regime and address both of these issues.

Table\,\ref{tab:best_fits} shows the best-fit double power-law parameters for three baryonic components. In Fig.\,\ref{fig:scaling_rels}, we show in blue dashed curves the 16$^{th}$ and 84$^{th}$ percentile region of the final best-fit curve adopted. The posterior sampled curves show the scatter as predicted by the bayesian analysis with \textsc{brms}. The best-fit curve parameters for the XPHM relation is modelled as the double power-law that fits best to red curve shown in \ref{fig:scaling_rels}(c). 

\begin{figure*}
  \includegraphics[scale=0.34]{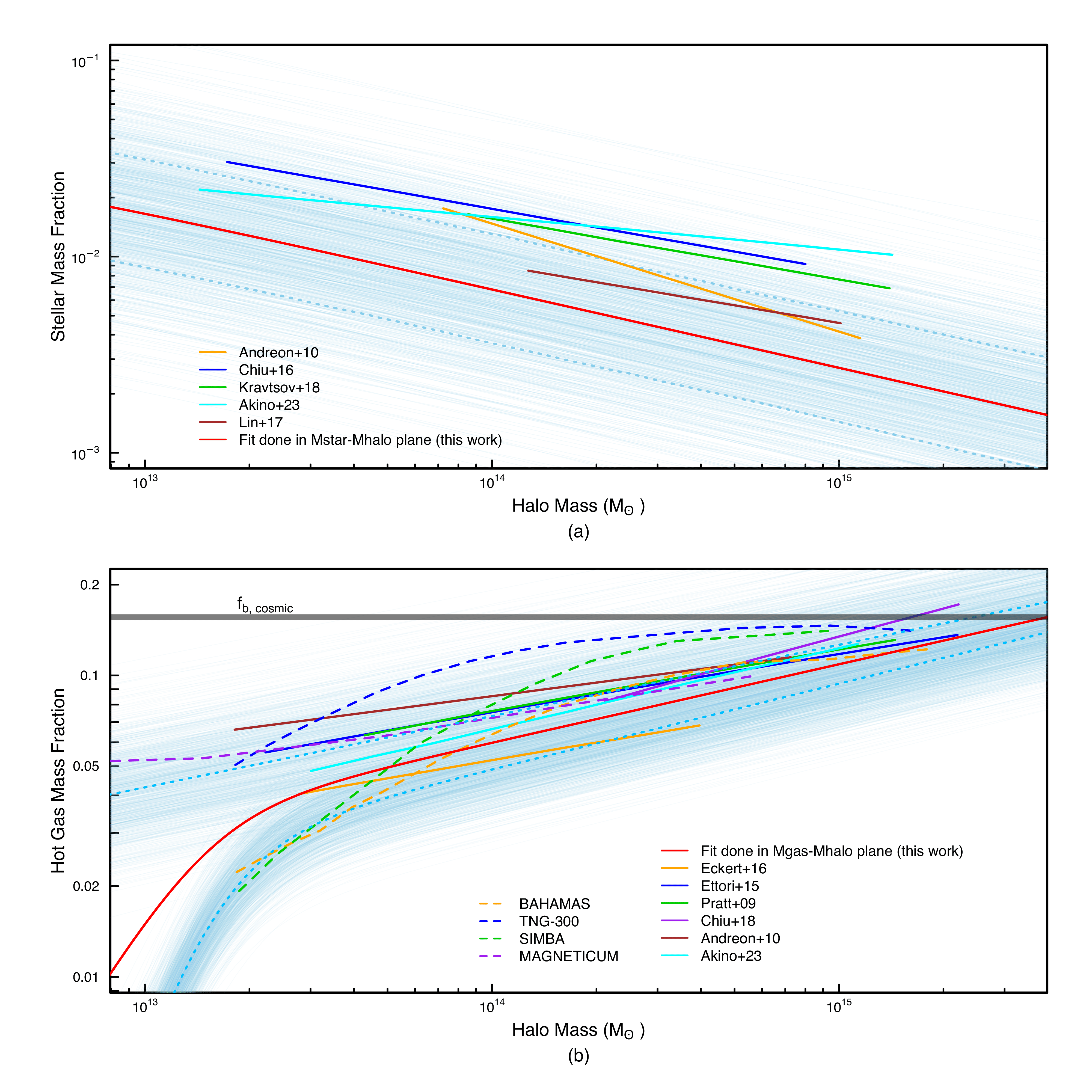}
 \caption{Comparison of baryon fraction - halo mass relations from this work with other literature measurements. (a) Stellar mass fraction - halo mass relation. Red line is our best-fit relation adapted from Fig. \ref{fig:scaling_rels}(a). Other literature scaling relations are from \citet{Andreon2010, Chiu2016, Kravtsov2018, Akino2022, Lin2017}. (b) Hot gas mass fraction - halo mass relation. Red line is our best-fit relation adapted from Fig. \ref{fig:scaling_rels}(c). Other literature scaling relations are observational results from \citet{Pratt2009, Andreon2010, Ettori2015, Eckert2016, Chiu2018, Akino2022} and hydrodynamical simulations (BAHAMAS-\citet{Mccarthy2009}, TNG300-\citet{Springel2018}, SIMBA-\citet{Dylan2020} and MAGNETICUM-\citet{Marini2024}). Blue solid curves in both the panels are the posterior sampled curves from the bayesian fitting which shows the scatter, and the dotted blue curves represent the 16th and 84th percentile of the scatter.}
 \label{fig:comparisons}
\end{figure*}

\subsection{Comparison to literature scaling relations}
Fig.\,\ref{fig:comparisons} shows the baryon fraction comparisons  derived in this work and other literature studies. To make all the comparisons consistent, we scale up all measurements within $R_{500}$ to $R_{200}$ using the scaling relation in \citet{Ragagnin2021}. In Fig.\,\ref{fig:comparisons}(a), we show the stellar fraction as a function of halo mass. We compare our result taken from the best-fit stellar-to-halo mass relation (Fig.\,\ref{fig:scaling_rels}(a)) to studies in high mass groups and clusters reported in the literature (\citealt{Andreon2010, Chiu2016, Kravtsov2018, Chiu2018, Akino2022, Lin2017}). Where necessary, stellar masses have been corrected for different IMF assumptions to a common Chabrier IMF (see conversions in \citealt{Madau&Dickinson2014}). 

Overall, our best-fit stellar baryon fraction is 2-3 times lower than other measurements at the high mass end shown in Fig.\,\ref{fig:comparisons}(a). This could be because the halos included in the above-mentioned studies may be sampling the upper end of our scaling relation, whereas our much larger sample of SDSS and GAMA halos show large intrinsic scatter. The above mentioned studies are based on X-ray selected sample excluding \citet{Andreon2010} and \citet{Lin2017}, which is different from our sample. The offset of the \citet{Andreon2010} can be explained based on the stellar mass estimation method. They obtain their stellar masses from luminosity adopting the M/L value derived by \citet{Cappellari2006} which however is biased higher as it is constructed solely of elliptical and lenticular galaxies, and those are dynamical M/L being sensitive to dark matter fraction \citep{Leauthaud2012}. \citet{Leauthaud2012} derived the stellar mass fraction ($\text{f}_{*}$) using HOD model as well as COSMOS X- ray group catalog and obtained a stellar fraction measurement similar to our results. 
%

In Fig. \ref{fig:comparisons}(b), we show the hot gas fraction comparison between our best-fit relation (Fig. \ref{fig:scaling_rels}(c)) and other scaling relations from observations (\citealt{Pratt2009, Andreon2010, Ettori2015, Eckert2016, Chiu2018, Akino2022}) and hydrodynamic simulations - BAHAMS, TNG300, SIMBA and MAGNETICUM (\citealt{Mccarthy2009, Springel2018, Dylan2020, Marini2024}). The two sets of blue curves are the posteriors of the fits without (single power-law) and with (double power-law) eRASS1 data. Our derived best-fit hot gas fractions, taken as the average of the two fits, is slightly lower than previous observational scaling relations by 1.1-2 times. This is mainly because our relation is driven by the new eRASS1 measurements which are arguably probing a more complete X-ray sample compared to previous relations which were primarily based on X-ray selected samples. The scaling relations from literature agree better with the upper dotted blue curve which is the fit to non-eRASS1 data. The offset could also be due to the different methods used to obtain the hot gas mass and halo mass. 

\begin{figure*}
 \includegraphics[scale=0.4]{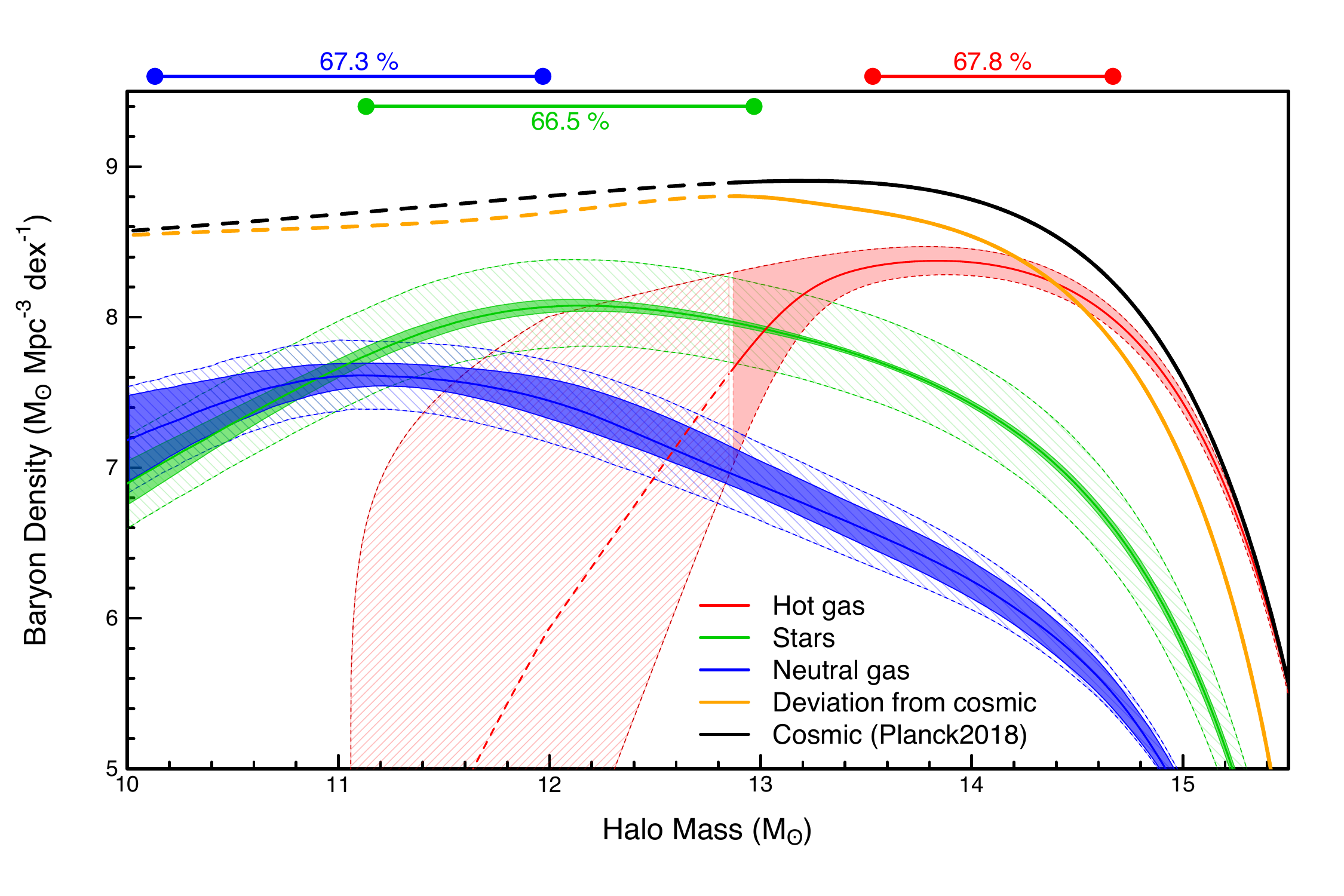}
 \caption{Baryon density for the three components - Neutral gas (in blue), stars (in green) and hot gas (in red). The baryonic scaling relations in Fig. \ref{fig:scaling_rels} have been multiplied by the Watson-HMF to obtain the following baryon density plot. This shows us the mass-density  contribution of each of baryonic component at different halo masses. As horizontal bands on top, we show the halo mass range which includes $\sim 67\%$ of the component baryon densities. In black, we show the baryon densities that should be present if all halos contain cosmic baryon fraction. In orange, we show the difference between the black and the sum of three baryon components. The shaded region shows the 1-$\sigma$ uncertainty on the mean baryon density curve for each component, while the hashed region shows the 16-84 percentile region of the total scatter in the relation.}
 \label{fig:combined_rho}
\end{figure*}

\begin{table*}
\centering
\renewcommand{\arraystretch}{1.5}
\begin{tabular}{ccccc} 
\toprule
            & $\text{GSR-HMF}$         
            & $\text{Planck-HMF}$         
            & $\text{Watson-HMF}$          
            & $\text{Literature}$      \\
\midrule
\textbf{$\Omega_\text{star}(\times 10^{-3})$} 
                 &    $2.71_{-0.96}^{+1.69}$
                 &    $3.4_{-0.41}^{+0.53}$
                 &    $2.09_{-0.18}^{+0.21}$ 
                 &    $2.18 \pm 0.03 $ \citep{Driver_GAMA_2022}    
                      \\

\textbf{$\Omega_\text{\HI}(\times 10^{-3})$}  
                 &   $0.45_{-0.25}^{+0.80}$
                 &   $0.71_{-0.17}^{+0.32}$
                 &   $0.49_{-0.12}^{+0.25}$
                 &   $0.38 \pm 0.04 \pm 0.04$ \citep{Zwaan2005} 
                 \\

\textbf{$\Omega_\text{neutral gas}(\times 10^{-3})$}    
                 &   $0.57_{-0.28}^{+1.84}$
                 &   $1.05_{-0.76}^{+1.64}$
                 &   $0.71_{-0.18}^{+0.39}$   
                 &   $0.60 \pm 0.09$ \citep{Zwaan2005, Fletcher2021} \\                
\textbf{$\Omega_\text{hot gas}(\times 10^{-3})$} 
                &	$5.77_{-1.31}^{+3.13}$
                &   $5.84_{-1.36}^{+3.73}$ 
                &   $2.58_{-0.66}^{+2.1}$ 
                &   $2.6 \pm 1.8$ \citep{Fukugita1998}	
\\ 
\bottomrule
\end{tabular}
\caption{ Baryon densities obtained in this work using baryon scaling relations combined with different HMF. GSR-HMF uses \citet{Driver_hmf_2022} without any constraints, Planck-HMF uses \citet{Driver_hmf_2022} with Planck18 $\Omega_\text{M}$ constraint, and Watson-HMF is based on \citet{Watson2013} with Planck18 cosmology with masses defined based on spherical overdensity definitions within $R_{200}$. Final column gives density estimates from literature calculated using other methods such GSMF \citep{Driver_GAMA_2022} or HIMF \citep{Zwaan2005} or $\text{H}_2$MF \citep{Fletcher2021}. }
\label{tab:omega}
\end{table*}

\section{The baryon densities versus halo mass}
\textcolor{black}{To convert our scaling relations to baryon-density halo-mass relations, we combine (multiply) the best-fit scaling relations with an adopted halo mass function (HMF). We then integrate across the halo mass range to obtain the total baryon density,}
\begin{equation}  \label{eq2}
\begin{split}
\rho_x &= \int_{M_\text{h,low}}^{M_\text{h,up}} \rho_x(M_\text{h}) dM_\text{h} \\
        &= \int_{M_\text{h,low}}^{M_\text{h,up}} \text{M}_x(M_\text{h}) \times \phi(M_\text{h}) 
 dM_\text{h},
\end{split}
\end{equation}
where $x$ can be stellar / \HI\ / hot gas component and $\phi$ is the HMF.


\textcolor{black}{A variety of HMFs are available in literature. In this work, we will calculate our baryon densities with two types of HMF - a theoretical one constructed from numerical simulations and empirical one with observational constraints. For the theoretical HMF, we use the \citet{Watson2013} based model (hereafter referred to as Watson-HMF). This is constructed from large N-body simulations with a range of box sizes ($\sim 11$ 
 Mpc to $\sim 6$ Gpc) and particle numbers ($3072^3$ to $6000^3$). For the Watson-HMF, we use the HMF constructed based on the spherical overdensity criterion ($\Delta_c=200$) and redshift $z=0$. In Fig. \ref{fig:combined_rho}, we show the baryon density curves obtained using Watson-HMF. }

\textcolor{black}{For the empirically derived form, we use the halo mass function from \citet{Driver_hmf_2022}. This was determined using a combination of GAMA, SDSS and REFLEXII data, extending the empirically studied HMF down to $10^{12.7} \text{M}_\odot$. At lower halo masses ($<10^{12.7} \text{M}_\odot$) we must resort to extrapolations of the empirical HMF. For this work we will adopt two forms of the empirical HMFs which bracket the range of possibilities. Firstly, the best-fit HMF which is constrained by observations down to $10^{12.7} \text{M}_\odot$, below which it is a power-law extrapolation. This will be referred to as the GSR-HMF. Secondly, a refitting of the HMF which is also constrained by observations down to $10^{12.7} \text{M}_\odot$ but now includes an additional constraint (prior) that the total integral of the HMF, across the entire halo mass range, should recover the matter density value from \citealt{Planck2018} ($\Omega_\text{M} = 0.31$). This will be referred to as the Planck-HMF.  In Table\,\ref{tab:omega}, along with Watson-HMF, we also report the values for the baryon densities calculated with both the GSR-HMF and the Planck-HMF.}

\textcolor{black}{The Watson-HMF is generated using the \textsc{hmf} Python code \citep{Murray2013}. For the GSR-HMF and Planck-HMF, we use the functional form and the parameters provided in \citet{Driver_hmf_2022}. In the main text of this paper, the plots and results are based on the Watson-HMF. The corresponding plots calculated using the empirical one, both GSR-HMF and Planck-HMF, are provided in the Appendix. }


\textcolor{black}{We estimate the uncertainties in our baryon densities by combining both the uncertainties in the HMF (wherever available) and the best-fit errors of the individual scaling relations. The Watson-HMF, as constructed from numerical simulations, does not have any uncertainties. Hence, the uncertainties in baryon density estimates using Watson-HMF come from the uncertainties in the best-fit of the individual scaling relations. The 1-$\sigma$ uncertainty on the mean baryon density curve for neutral gas, stellar and hot gas components are shown by the green, red and black shaded regions.  The hashed regions shows the 16$^{th}$ and 84$^{th}$ percentile region of the total scatter in the baryon density, which includes the intrinsic scatter in the scaling relations as well. }

On the other hand, uncertainties for the empirical HMFs are taken from \citet{Driver_hmf_2022}. The errors are maximally included by combining HMFs sampled from the covariance matrix from \citet{Driver_hmf_2022} with the posterior samples generated from the bayesian regression for each of the scaling relations. The associated baryon density relation using GSR-HMF is shown in  Fig.\,\ref{fig:combined_rho_gsr}. The HMF errors are noticeably larger for the GSR-HMF as compared to the Planck-HMF (\citealt{Driver_hmf_2022}). This explains the larger spread of 16$^{th}$ and 84$^{th}$ percentile regions of the baryon densities estimated using the GSR-HMF (Fig. \ref{fig:combined_rho_gsr}) compared to those calculated using the Planck-HMF (Fig. \ref{fig:combined_rho_planck}), as well as the larger uncertainties in the GSR-HMF derived baryon densities compared to Planck-HMF as shown in Table\,\ref{tab:omega}. 

\textcolor{black}{Finally, to obtain the cosmic baryon density for an individual baryonic component, we integrate the baryon density - halo mass curve across the entire halo mass range using Eq. \ref{eq2} and normalize with the critical density of the universe ($\Omega_x = \rho_x/\rho_\text{c}$). These are shown in Table\,\ref{tab:omega}. The values under the column Watson-HMF are calculated by integrating the curves in Fig. \ref{fig:combined_rho}. The corresponding values under GSR-HMF and Planck-HMF are based on the Fig. \ref{fig:combined_rho_gsr} and Fig. \ref{fig:combined_rho_planck} respectively. }

\subsection{Stellar Component}
The stellar-halo mass relation for our data is shown in Fig.\,\ref{fig:scaling_rels}(a). The best-fit line shows the change in slope of the relation around $\sim 10^{11.72} \text{M}_\odot$ which has been noted previously in several studies (\citealt{Leauthaud2012, Eckert2017, Hutchens2023}).\textcolor{black}{ This is interpreted as a transition scale where halos move from isolated to group regime.} The stellar mass density as a function of halo mass is shown by the green curve in Fig.\,\ref{fig:combined_rho}. We see that the dominant contribution to the stellar mass density arises from the halo-mass range $10^{11.1}-10^{13} \text{M}_{\odot}$. Hence most stars reside in galaxies within dark-matter halos in this mass range. The cosmic stellar baryon density comes out to be $\Omega_{\text{star}}=2.09^{+0.21}_{-0.18} \times 10^{-3}$. From this, $\sim 60\%$ of the stellar mass density lies in halos with masses above $> 10^{12} \text{M}_\odot$.

We can compare our total stellar matter density value with that obtained from integrating the GSMF. \citet{Driver_GAMA_2022} obtained a $\Omega_\text{star}$ value of $(2.18 \pm 0.03) \times 10^{-3}$, which is consistent with our result. They used the GAMA data below $z<0.1$ to calculate the stellar mass density. \textcolor{black}{ \citet{Rodriguez-Puebla2020} estimated the cosmic stellar mass density from GSMF constructed using SDSS data. They obtain a value of $\Omega_\text{star}$ value of $(20.40 \pm 0.08) \times 10^{-4}$, which is also consistent with our result. } The GSMF method has smaller errors, so is more reliable but nevertheless provides a basic confirmation of our methodology to get the cosmic stellar mass density. 

\subsection{\texorpdfstring{Neutral Gas Component (\HI\ +$\text{He}$ +$\text{H}_2$)}{Neutral Gas Component}}
The \HI-halo mass relation for our data is shown in Fig. \ref{fig:scaling_rels}(b). The best fit line shows a change in slope of the relation around $\sim 10^{11.1} \text{M}_\odot$. \textcolor{black}{This scale has been reported as a halo mass threshold scale, below which centrals are mostly gas-dominated disks which transitions to gas-poor systems at higher halo masses \citep{Kannappan2008, Kannappan2013, Eckert2017, Hutchens2023}.}  We note that while it is standard in the radio community to report the neutral gas scaling relation in terms of \HI\ versus halo mass, we now wish to include the H$_2$ and $He$ components. To include the contributions of He and heavier
elements, we now multiply the \HI\ masses by the standard factor of 1.36
(\citealt{Saintoge&Catinella2021}, \citealt{Guo2023}). We include the molecular gas ($\text{H}_2$) contribution using the empirical model - NeutralUniverseMachine \citep{Guo2023}, which is built upon the UniverseMachine model \citep{Behroozi2019} and self-consistently models the dark matter halos and star-formation rate of individual galaxies. In this model, the $\text{H}_2$ content in dark matter halos is modelled using a functional form which depends on stellar mass, redshift and offset from the star formation main sequence. The $\text{H}_2$ model is constrained using observations of the $\text{H}_2$ mass function, $\text{H}_2$-stellar mass relation, $\text{H}_2$-\HI\ mass ratio for $z\sim0$ and molecular gas density measurements. From here on, we combine the 1.36\HI\ (including contributions from He and other elements) and $\text{H}_2$ baryon contents and describe them together as 'Neutral gas'.

We obtain a cosmic \HI\ density  of  $\Omega_\text{\HI} = (0.49^{+0.25}_{-0.12}) \times 10^{-3}$. This value is consistent with the estimates from more direct $\Omega_\text{\HI}$ studies. \citet{Zwaan2005} obtained a $\Omega_\text{\HI} = (0.38 \pm 0.04 \pm 0.04) \times 10^{-3}$ for $z<0.06$ using \HI\ Parkes All Sky Survey (HIPASS) data. Similar values for $\Omega_\text{\HI}$ have also been obtained in other studies \citep{Rodriguez-Puebla2020, Rhee2023}. The consistency between our result and previous $\Omega_\text{\HI}$ measurements, once again validates our method for estimating the cosmic baryon density.  

The neutral gas mass density as a function of halo mass is shown by the blue curve in Fig.\,\ref{fig:combined_rho}. Including the $\text{H}_2$ contribution from \citet{Guo2023} and correcting for He and heavier elements, we obtain a cosmic neutral gas density of $\Omega_\text{neutral gas} = (0.71^{+0.39}_{-0.18}) \times 10^{-3}$. \citet{Fletcher2021} determined the cosmic abundance of $\text{H}_2$ using the xCOLD GASS (eXtended CO Legacy Database for GASS; \citealt{Saintoge2017}) survey to be $\Omega_{\text{H}_2} = (7.62 \pm 0.67) \times 10^{-5}$. We can combine this $\text{H}_2$ abundance measurement with the \HI\ abundance measurement from \citet{Zwaan2005} to estimate the cosmic neutral gas abundance from literature. This neutral gas estimate is consistent with our $\Omega_\text{neutral gas}$ after accounting for the $He$. \textcolor{black}{ \citet{Rodriguez-Puebla2020} used conditional probability distribution functions from \citet{Calette2018} and obtained a cosmic neutral gas density estimate of $\Omega_{\text{neutral gas}} = (6.85 \pm 0.92) \times 10^{-4}$, which is also consistent with our results.}

\subsection{X-ray Component}
The X-ray detected plasma-halo mass relation is shown in Fig. \ref{fig:scaling_rels}(c) and the X-ray mass density as a function of halo mass is shown by the red curve in Fig.\,\ref{fig:combined_rho}. We can see that the slopes of the eRASS1 and non-eRASS1 dataset (the upper and lower bound of the red shaded region) are similar at the cluster regime and this is where the scatter in the relation is the smallest. The largest discrepancy is at the group scale especially at  the low mass end ($< 10^{13.3} \text{M}_\odot$) where CGM stacking results show a consistency with the single power-law extrapolation, but the individual eRASS1 detections point towards a turnover. Hence the large uncertainty in the hot gas density at the low mass group regime. 

The cosmic hot gas density estimate us $\Omega_{\text{hot gas}}=2.58^{+2.1}_{-0.66} \times 10^{-3}$. This is a first direct estimate of the cosmic hot gas density ($\Omega_\text{hot gas}$) as a function of halo mass which includes empirical constraints down to such low halo masses which has become possible because of the large sample from the eROSITA survey. Our estimate seems consistent with the values reported in \citet{Fukugita&Peebles2004}.

\section{Discussion}
\begin{figure*}
 \includegraphics[scale=0.4]{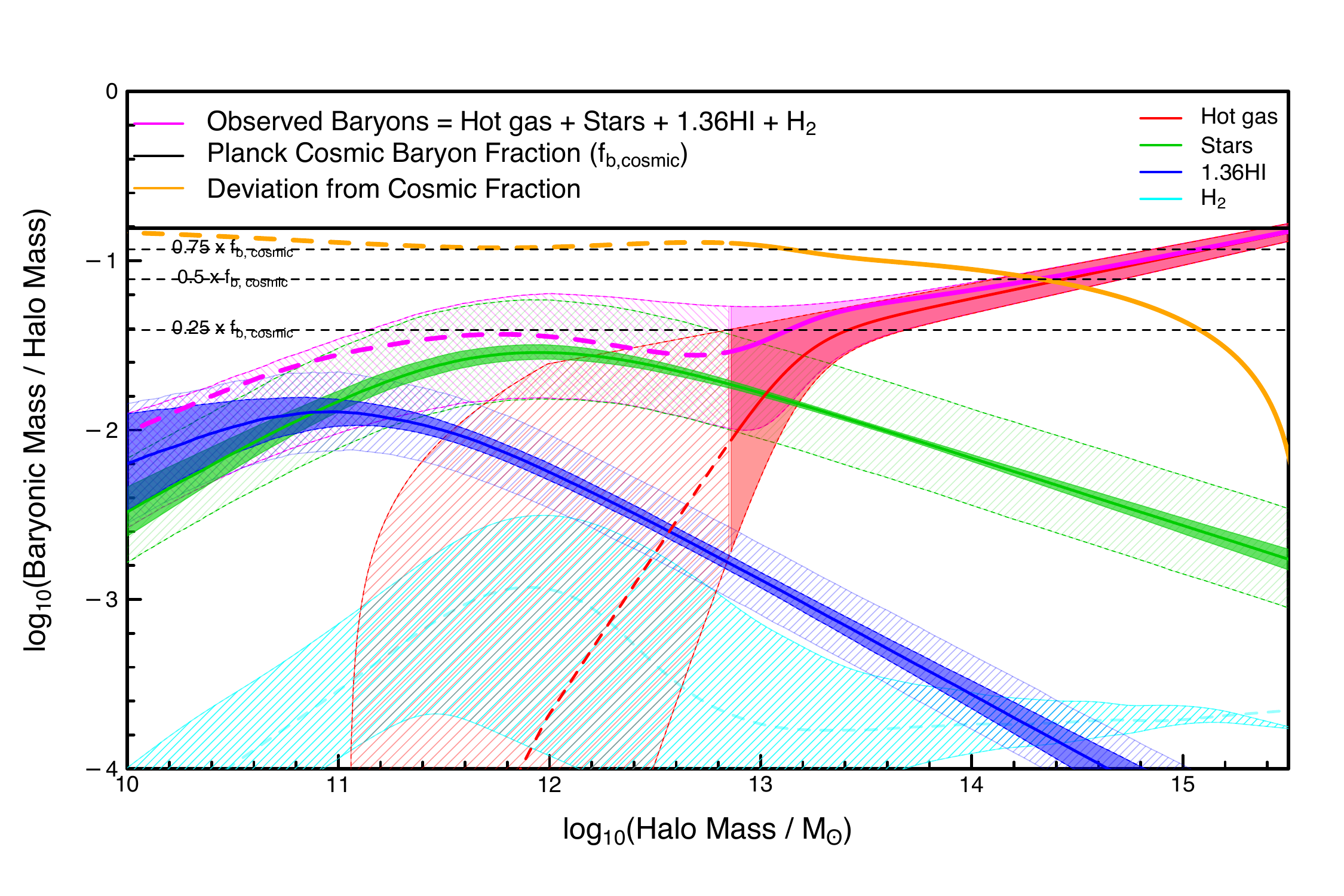 }
 \caption{Baryon fraction, $f_\text{b}=M_\text{b}/M_\text{h}$, as a function of halo mass for $\text{H}_2$ (in cyan), \HI\ + He (in blue), stars (in green) and hot gas (in red). The shaded region represents the best-fit errors, whereas the hashed region shows the 16-84 percentile confidence interval for the total scatter in the relations for each of the components. The horizontal black line shows the cosmic baryon fraction ($f_\text{b, cosmic} = 0.156$) based on \citet{Planck2018} measurements. In magenta, the sum of the three baryonic components are shown. Also shown is the deviation of the total observed baryons from the cosmic fraction in orange. }
 \label{fig:baryon_fraction}
\end{figure*}

We combine the scaling relations for our three baryonic components (Hot gas, Stars and Neutral gas) and now look at the individual and combined baryon fractions as a function of halo mass. This is shown in Fig.\,\ref{fig:baryon_fraction} where we plot the mass fraction given by, $f_{\text{b}}=M_\text{b}/M_\text{h}$,  as a function of halo mass for the \HI\ (in blue), stars (in green) and hot gas (in red). To include the contributions of He and heavier elements, we add their contributions to \HI\ by multiplying by $\times 1.36$ (as indicated in the key), and we additionally plot the $\text{H}_2$ fraction in cyan.  Stellar and \HI\ fractions are empirical down to $10^{10} \text{ M}_\odot$, whereas X-ray fractions are available only down to $10^{12.8} \text{ M}_\odot$ in halo mass with any high degree of certainty. 

At the high halo mass end, where measurements are robust, the dominant source of baryonic mass is the hot gas, this transitions to stellar-mass as the dominant baryonic component (of the detected components) at a halo-mass of $\sim  10^{13} \text{ M}_\odot$. Below $\sim  10^{11} \text{ M}_\odot$ , we notice the neutral gas contribution starting to dominate but with significantly broad error distributions and overlap with the stellar curve.  


In magenta, we show the total baryon fraction obtained by summing the contributions of all three components. It can be seen that except at the highest halo mass end ($\sim 10^{15} \text{ M}_\odot$), the total baryon fraction falls short of the expected universal baryon fraction value of 0.156 obtained from CMB studies \citep{Planck2018}, with the observed baryon fraction falling below 50\% for halos with mass ($< 10^{14} \text{ M}_\odot$). For the halo mass range of $< 10^{13}-10^{14} \text{ M}_\odot$, we see almost consistent baryon fraction of 25\% of $f_\text{b, cosmic}$. Similar decreasing baryon fraction trends have been seen in previous studies which generally looked at the contribution of hot gas and stars for halo masses above $10^{13} \text{ M}_\odot$ \citep{Gonzalez2013, Akino2022}. Our total baryon fraction measurements are largely in agreement with the results from contemporary cosmological hydrodynamical simulations, although the exact trend has a large scatter depending on the model (\citealt{Oppenheimer2021}).

In Fig.\,\ref{fig:combined_rho}, we show the actual and combined baryon densities as a function of halo mass for the different baryonic components. The stellar mass density peaks at a halo mass of $10^{12} \text{ M}_\odot$ and $\sim 67\%$ of the stellar component lies between $10^{11.4}-10^{13.6} \text{ M}_\odot$. The neutral gas mass density peaks at a halo mass of $10^{11} \text{ M}_\odot$ and $\sim 67\%$ of the neutral gas component lies between $10^{10.8}-10^{12.4} \text{ M}_\odot$. The hot gas mass density peaks at a halo mass of $10^{14.05} \text{ M}_\odot$ and $\sim 68\%$ of the hot gas component lies between $10^{13.6}-10^{14.8} \text{ M}_\odot$. 

On Fig.\,\ref{fig:combined_rho} and {\ref{fig:baryon_fraction} we show in orange the implied missing baryon component which we will discuss in Section\,4.1.

Fig.\,\ref{fig:combined_omega} shows the cumulative contribution (by integrating from high mass to low mass) of the individual and total baryon density as a fraction of the Planck2018 baryon density ($\Omega_\text{b} = 4.6 \times 10^{-2}$). It is obtained by integrating the curves in Fig. \ref{fig:combined_rho}, which shows the baryon density as a function of halo mass for all three components. The red, green and blue shows the contribution of baryons in the form of hot gas, stars and neutral gas, respectively. In black, we show the total which is the Planck2018 cosmic baryon fraction converted into total baryon density using the HMF. Subtracting the observed baryons from the total shows the missing component which is shown in orange.  Combining the three components - the stars, neutral gas and hot gas, we can account for at max 20\% of the cosmic baryon density in halos with mass greater than $10^{10} \text{ M}_\odot$ at $z=0$.

\begin{figure*}
 \includegraphics[scale=0.4]{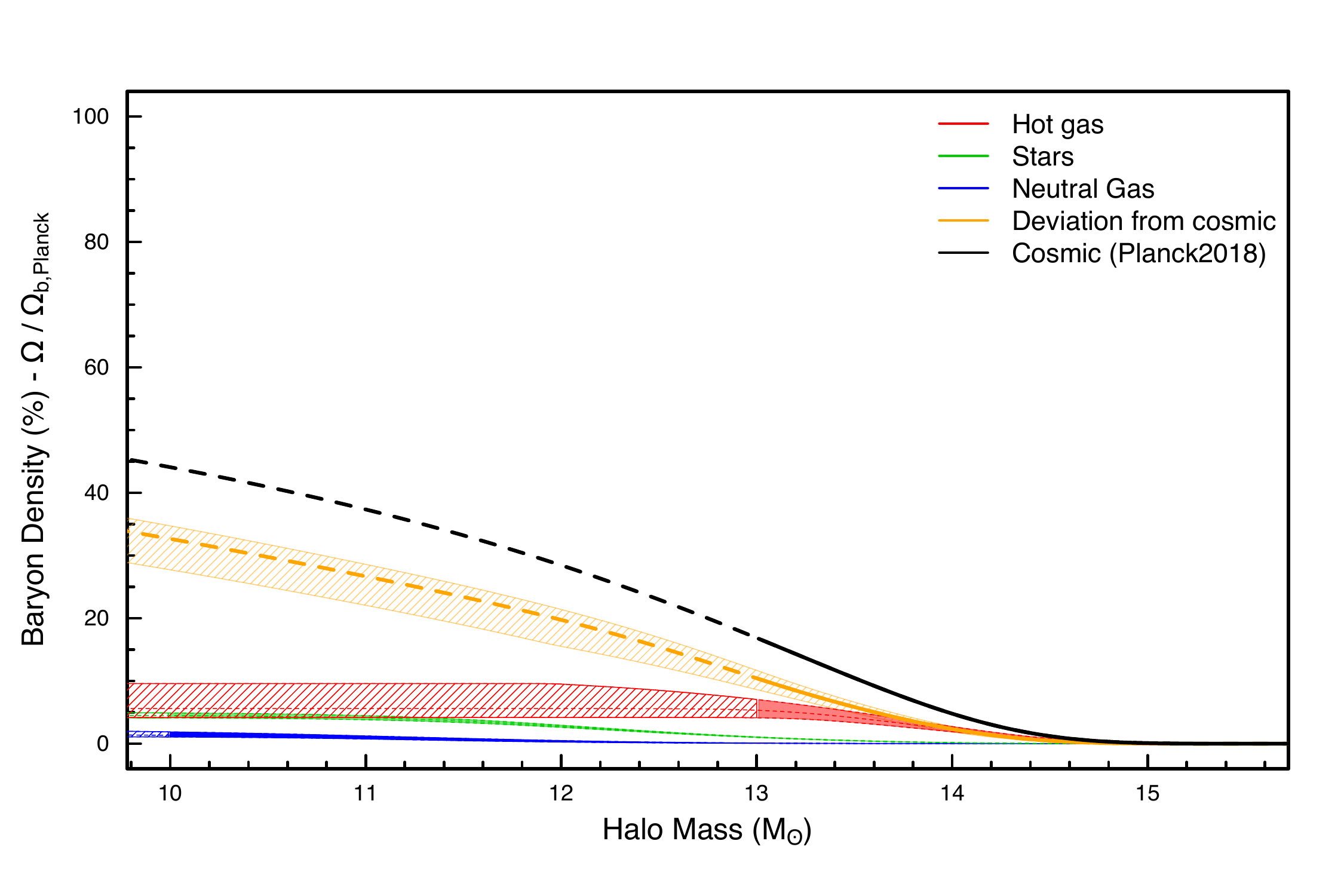}
 \caption{Cumulative contribution of the cosmic baryon density, by integrating from the highest to lowest halo mass, shown as a fraction of the Planck2018 cosmic baryon density value. Cosmic baryon density for the three components - Neutral gas (in blue), stars (in green) and hot gas (in red) are shown. The cyan line shows the missing baryon component needed to match the to the cosmic baryon fraction at each halo mass. The black shows the sum of all the observed and missing baryonic components. }
 \label{fig:combined_omega}
\end{figure*}

\subsection{Missing Mass}
Observational studies to probe the elusive WHIM have been reported using observational techniques such as SZ-CMB studies (\citealt{Planck2013, deGraff2019, Tanimura2019, Das2019, Singari2020, Tanimura2022}), X-ray absorption line studies (\citealt{Werk2014, 
Nicastro2018, Nicastro2023, Mathur2023}) and Fast Radio Bursts (FRBs; \citealt{Lorimer2007, Macquart2020, Yang2022, Simha2023}). These have provided us with  methods to estimate the different phases of the total ionised baryon content, and are providing fresh constraints on the total baryon density. But a complete census of the various ionised gas phases across a wide halo mass range is still lacking. 

One possible reason for the discrepancy between the total and observed baryons, shown by the orange curve in Fig. \ref{fig:baryon_fraction}, could be that remaining baryons inside halos may be in the form of warm ionized gas ($\text{T} \sim 10^{5-6}\text{K}$). \citet{Lim2017b} detected kinetic Sunyaev-Zel'dovich effect (kSZE) signals by performing stacking of $\sim 40\,000$ galaxy groups across a halo mass range of $12.3 \le \text{log}_{10}(\text{M}_{200c}/\text{M}_\odot) \le 14.7$, using Planck temperature maps and the SDSS-based group catalogue \citep{Yang2007}. Using the ionized gas measurements from the kSZE, they obtain a baryon fraction consistent with the Planck cosmic baryon fraction across a wide halo mass range $10^{12.3}-10^{14.7} \text{M}_\odot$. This supports the idea that all the missing baryons within halos in our study are present in the form of a warm-hot intergalactic medium (WHIM). \citet{Eckert2017} found that the group baryonic mass functions are shifted lower by a factor of $\sim 2$ compared to the cosmic baryon fraction, but can be reconciled by including WHIM estimates of 40-50\%. Similar kSZ measurements in galaxy-scale halos have shown that the baryon measurements are consistent with the cosmic baryon fraction \citep{Hernandez2015}. Apart from WHIM, the cool ionised gas phase (T $\sim 10^4 \text{ K}$) is also not included in our halo baryon budget. These can be detected by tracers such as H$\alpha$ emission lines or absorption measurements from far-UV and Ly$\alpha$ \citep{Prochaska2011, Thom2012}. These could have significant contributions in the groups or further lower halo mass regime \citep{Prochaska2013, Zhang2020}. Future works including these remaining baryon components would bring us closer to a complete baryon inventory in halos. 

Another consideration is whether some or all of the remaining baryons are located at a larger radius from the halo centre, i.e. beyond $\text{M}_{200}$. These baryons could have either been expelled due to various feedback mechanisms from within the halos or alternatively may never have been accreted into halos. Simulations provide the ideal scenario to test these cases, and both the EAGLE and IllustrisTNG models have seen evidence of this \citep{Ayromlou2022, Mitchell2022}. \citet{Ayromlou2022} looked at the baryon distribution in halos as a function of virial radius for three sets of cosmological hydrodynamical simulations - IllustrisTNG (TNG50, TNG100 and TNG300), EAGLE and SIMBA - across a halo mass rage of $10 \le \text{log}_{10}(\text{M}_{200c}/\text{M}_\odot) \le 15$. 

In all the simulations, they observed a baryon fraction within the halo ($\le R_{200c}$) to be lower than the cosmic baryon fraction, for all halos with halo masses below $10^{14} \text{M}_\odot$, which is consistent with our results. These baryons are however recovered when  traced out to large virial radii ranging from 4-13 $R_\text{200c}$ depending on the different models and halo mass (see Fig. 2 and 7 of \citealt{Ayromlou2022}). The reason for the baryons to be located in the halo outskirts is attributed to stellar and AGN feedback. \citet{Mitchell2022} used EAGLE simulations to study the baryon mass budgets in halos ranging from $10 \le \text{log}_{10}(\text{M}_{200c}/\text{M}_\odot) \le 14.5$. In all the EAGLE halos nearly 50-80\% of the baryonic mass accreted, seemed to be ejected outside the halos by $z=0$, with the ejected fraction increasing towards lower halo masses (see Fig.\,6 of \citealt{Mitchell2022}). Although the exact values are model-dependent, both these studies indicate that the missing fraction of baryons that we are unable to account for could potentially be outside the halo at larger virial radii. 

Similar results have been found in other studies as well \citep{Galarraga2022}. \citet{Sokolowska2016} found that 20-30\% of the WHIM is pushed out to 1-3 virial radii in simulation studies of Milky-Way like halos. \citet{Wang2017} used Numerical Investigation of a Hundred Astrophysical Objects (NIHAO) galaxy formation simulations and reported that all their halos (from dwarfs to Milky-Way) are deficient in baryons which can be relocated beyond twice the virial radius. However, all these simulation results are inconsistent with the kSZE predictions of \citet{Lim2017b}. Future observational studies focussing on WHIM as well as probing baryons out to larger radii will help in confirmation of either of these scenarios. 

\subsection{Biases in the study}

\subsubsection{Choice of HMF}
\textcolor{black}{ We have used the Watson-HMF to derive the results mentioned in Section\,3. The same analysis has been repeated with the empirical HMFs and the results are presented in the Appendix. In Fig. \ref{fig:hmf}, we provide a comparison between the \citet{Watson2013} and \citet{Driver_hmf_2022} used in this work.}

\textcolor{black}{ The uncertainties on the baryon densities derived using the theoretical HMF is lower than the ones with empirical HMFs. This is solely because the theoretical HMFs have no associated uncertainties and hence their baryon density uncertainties have only the contribution from the scaling relations. There are a wide variety of other theoretical HMFs constructed using numerical simulations in the literature \citep{Reed2007, Tinker2008, Bhattacharya2011, Angulo2012, Ishiyama2015}. Each of them vary from each other slightly in the slope or normalization either in the low or high-mass end. The HMFs also vary in their shape depending on how the extent of the halo is defined (spherical overdensity or friends-of-friends halos), which themselves can contain different models.  Hence, the baryon densities derived using a different HMF will yield slightly different results. }
\textcolor{black}{ While the theoretical HMFs are based on numerical simulations, the empirical ones are based on optical and X-ray group/cluster catalogues. These could have biases arising from halo mass uncertainties, incompleteness, cosmic variance and the accuracy of Eddington bias correction, which could impact the final form of the HMF. In between the two empirical HMFs used in this work, uncertainties on the baryon densities derived using the Planck-HMF are smaller compared to the ones derived from GSR-HMF. This is due to the shape of the Planck-HMF being strongly constrained both at the high-mass end by the data and at the low-mass end by the $\Omega_\text{M}$ constraint. The GSR-HMF integrates to give a $\Omega_\text{M}$ of $\sim$ 66\% of the Planck value (0.31) whereas the Planck-HMF is constrained to integrate to yield the precise Planck $\Omega_\text{M}$ value (see Fig.\,13 of \citealt{Driver_hmf_2022}). In comparison, the Watson-HMF integrates to $\sim 75\%$ of $\Omega_\text{M}$. }
\textcolor{black}{Both the empirical HMFs are very similar upto $\sim 10^{12.7} \text{M}_\odot$, and differ significantly in the lower halo mass regime where empirical constraints on the HMF are not currently available. The Planck-HMF has the intrinsic assumption that all of the matter is attached to haloes of some mass, which is true according to the HaloModel \citep{Cooray&Sheth2002}. However, this may not be true for the real Universe (see e.g., \citealt{Angulo&White2010, Wang2011}). \citet{Diemer2020} used N-body simulations to calculate halo mass functions with halo boundaries defined by splashback radius, and they report that only 50-60 \% of the dark matter particles reside within the halo. Hence, with the use of two empirical forms, GSR-HMF and Planck-HMF, we bracket the range of possibilities of the true HMF.}

\subsubsection{Heterogeneous sample}
Throughout this work, we have used a variety of datasets from the literature to construct our scaling relations. As we are trying to study a wide range of halo mass combined with a multi-wavelength approach (stellar, \HI\ , X-ray observations), the use of a heterogeneous sample at this time cannot be avoided. Due to this, there could be potential selection effects at different halo mass regimes which have not been accounted for (unknown unknowns). The various halo masses used in the study have been, wherever possible, those that most closely resemble a mass estimate within an aperture defined by an overdensity of 200 times the critical density of the Universe ($\text{M}_{200}$). However, halo mass estimation due to different tracers (e.g. using velocity dispersion, weak-lensing, rotation-curve modelling) are known to be biased, and hence could introduce additional sources of uncertainty.  

A future improvement, and one we aim for in the upcoming 4MOST WAVES survey \citep{Driver2019}, is to derive the \HI\, stellar and hot gas masses for the same halo sample with uniformly defined halo masses. While not possible now, this should be possible within the next decade. 

\section{Conclusions}

We construct baryonic mass scaling relations as a function of halo mass to describe the stellar, \HI\ and hot gas mass content in halos at $z=0$. The stellar mass and \HI\ mass relations empirically extend down to $10^{10} \text{ M}_\odot$, while the hot gas scaling relations is empirically constrained down to $10^{12.8} \text{ M}_\odot$. We fit double power-law models to each of the scaling relations. 

We then use our scaling relations, multiplied by a halo mass function to determine the baryon density distributions of stars, neutral gas and hot gas as a function of halo mass. Through our halo-based approach we estimate a cosmic baryon density in halos, $\Omega_\text{b} = 5.38^{+2.14}_{-0.71} \times 10^{-3}$, which includes the contributions of stars, neutral gas and hot gas ($>10^6 \text{K}$) components across the entire halo mass range using Watson-HMF at $z=0$. Out of  this, the neutral gas contributes $\sim 1$\%, stellar content contributes $\sim 5$\% and hot X-ray gas contributes $\sim 6$\% to the cosmic baryon density ($\Omega_\text{b, cosmic} = 4.6 \times 10^{-2}$, \citealt{Planck2018}), resulting in a total baryon content in the three components in bound halos in the local universe to be $\sim 12\%$.  

The total baryon density results using the empirical HMFs, GSR-HMF and Planck-HMF, are $\Omega_{\text{b}}^{\text{GSR-HMF}}  = 9.13^{+5.25}_{-2.02} \times 10^{-3}$ and $\Omega_{\text{b}}^{\text{Planck-HMF}} = 10.36^{+3.40}_{-2.18} \times 10^{-3}$ respectively. The cosmic baryon density estimates using the empirical HMF are comparatively higher, ranging from $19-23 \%$.

Our cosmic stellar and \HI\ density estimates using Watson-HMF are consistent with other literature measurements of these in the local universe. The results are also consistent with GSR-HMF and Planck-HMF albeit within their large uncertainties. Although there are other methods to estimate the individual baryon densities separately, our halo-based approach can be used to estimate the baryon densities consistently for various baryonic components.  

Assuming all the halos have a cosmic baryon fraction, we see that the observed baryons can account for less than 50\% of the baryon fraction except in the largest cluster halos ($M_\text{h}>10^{14.5} \text{ M}_\odot$). The major portion of the unaccounted baryons could most likely be in the form of WHIM. However, if a significant fraction of the baryons have been ejected outside, then the halos themselves do not need to be representative of the cosmic baryon fraction. This has been seen in numerical simulations, but needs confirmation by observational studies probing the baryon content in halos out to larger radii. Combining the current and upcoming large surveys with ASKAP, eROSITA and 4MOST, would provide us with the ideal multi-wavelength dataset that will help us in addressing several of these issues. However, probing the multi-phased ionised gas at different environments, which still remains challenging, will be necessary to have a complete census of all the baryonic components. 

\section*{Acknowledgements}
We thank the referees for their feedback that helped improve the paper. We thank Matt Owers and Oguzhan Cakir for useful discussions regarding SAMI-cluster data. We thank Hong Guo for providing data from the NeutralUniverseMachine model. We thank Zackary Hutchens for providing data from their work. We also thank Andrea Merloni, Claudia Lagos, Esra Bulbul, Emre Bahar, Kris Walker, Luke Davies, Luca Cortese, and Fabrizio Nicastro for useful discussions. AD acknowledges the support from University Postgraduate Award and Scholarship for International Research Fees from the University of Western Australia. SPD acknowledges financial support via an ARC Australian Laureate Fellowship award (FL220100191). DO is a recipient of an Australian Research Council Future Fellowship (FT190100083) funded by the Australian Government. PP has received funding from the European Research Council (ERC) under the European Union’s Horizon Europe research and innovation programme ERC CoG (Grant agreement No. 101045437).

\section*{Data Availability}

All the data used in this work are taken from previous published results tabulated in Table \ref{tab:data}, which have been described in Section 2.



\bibliographystyle{mnras}
\bibliography{example} 




\appendix

\section{HMF}

\begin{table*}
\centering
\begin{tabular}{lll}
\toprule
{\textbf{Stellar data}} & {\textbf{\HI\ data}} & {\textbf{X-ray data}}
                 \\
\midrule
 GAMA \citep{Robotham2011, Driver_GAMA_2022}      & \citet{Guo2020}       &  eRASS1 \citep{Bulbul2024}  \\
 
 SDSS \citep{Tempel2014, Chen2012}      &  \citet{Dev2023}       &  \citet{Mulroy2019}  \\
 
 SAMI-cluster \citep{Owers2017}      & \citet{Rhee2023}       &  \citet{Lovisari2015}  \\
 
 SPARC \citep{Lelli2016, Posti2019}      &      \citet{Hutchens2023}  &  \citet{Gonzalez2013}  \\
 
        &  \citet{Biswas2023}      &  \citet{Pearson2017}  \\
        
        &  SPARC \citep{Lelli2016, Posti2019}       &  \citet{Vikhlinin2006}  \\
        
        &       &  \citet{Arnaud2007}  \\
        &      &  \citet{Sun2009}  \\
                
\bottomrule
\end{tabular}
\caption{List of the various datasets used in the construction of the stellar, \HI\ and X-ray mass to halo mass scaling relations in Fig. \ref{fig:scaling_rels}. }
\label{tab:data}
\end{table*}

\begin{figure}
 \begin{minipage}{\linewidth}
  \includegraphics[width=\linewidth]{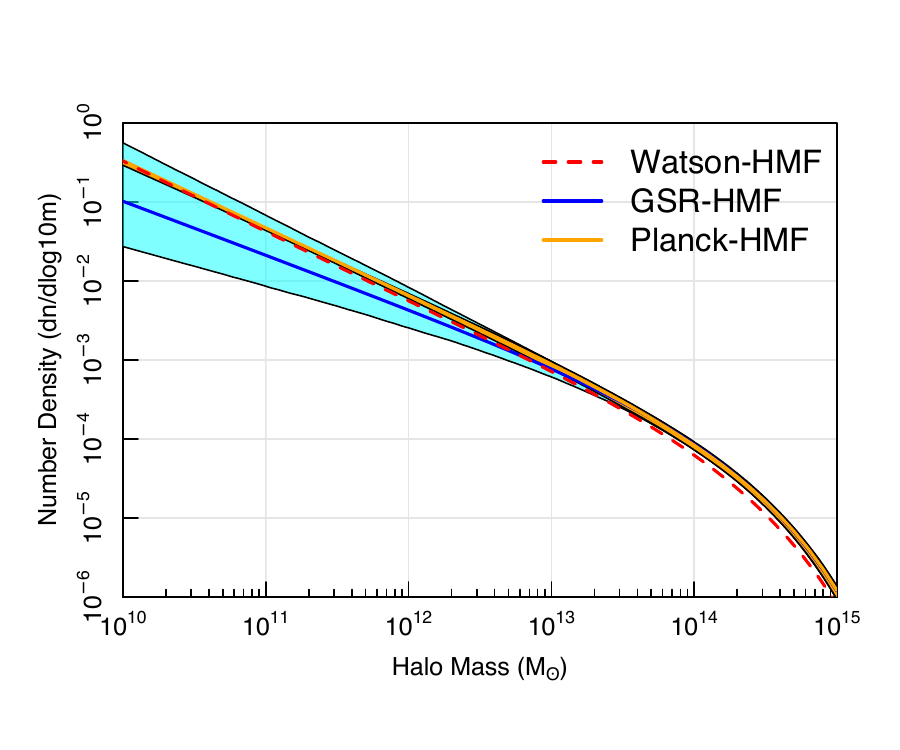}
  \caption{Comparison between the Watson-HMF and the two forms of empirical HMF. Watson-HMF, shown as red dotted lines, is based on numerical simulations \citep{Watson2013}. The two empirical HMFs are based on fitting GAMA, SDSS and REFLEX-II datasets \citep{Driver_hmf_2022}. The GSR-HMF (blue curve) is the empirical HMF with a power law extrapolation for low halo mass end, whereas the Planck-HMF (orange curve) has the \citet{Planck2018} $\Omega_\text{M}$ constraint. The cyan region shows the 16th and 84th percentile uncertainty region of the GSR-HMF. The uncertainty in the Planck-HMF is shown by the spread in the orange curve and is relatively quite small compared to uncertainty in GSR-HMF shown in cyan.}
  \label{fig:hmf}
 \end{minipage}
\end{figure}

\textcolor{black}{In this work, we use three different HMFs to calculate the baryon density. In Fig. \ref{fig:hmf}, we provide a comparison between the theoretical and empirical HMFs. At the high mass end $(M_\text{h} > 10^{13} \text{ M}_\odot)$, Watson-HMF is lower than the empirical HMFs. This is the main reason for the baryon density estimates at high-mass end based on empirical HMFs being higher compared to Watson-HMF as shown in Columns 2 and 3 of Table \ref{tab:omega}. The relative difference in the total cosmic baryon density for \HI\ and stellar component are comparatively less between GSR-HMF and Watson-HMF, as the two HMFs cross-over and Watson-HMF has a higher value at the low mass regime. In Fig. \ref{fig:hmf}, we can also see the uncertainties in the empirical HMF, especially GSR-HMF, being quite large and hence result in larger uncertainties in the corresponding baryon baryon density values. }

\textcolor{black}{The use of the Planck-HMF fundamentally assumes that all the mass in the universe is bound in halos within their virial radius ($\text{M}_{200}$) and the mass function is a smooth extension down to the lowest halo mass. However, this picture is incomplete as we know that matter is also present in the inter-galactic medium / inter-cluster medium outside halos. In simulations, this would be comparable with particles which are either unbound or do not end up within the virial radii of any halo. Hence, integrating an empirical HMF which takes into account only the masses inside the virial radius of a halo, should not necessarily integrate to give $\Omega_\text{M}$, but rather a lower value. Even if we switch to another HMF which is constructed based on all the bound mass within a halo rather than the mass within virial radius, this also should not be constrained to $\Omega_\text{M}$ because there could be diffuse matter unbound to halos. Hence, we prefer GSR-HMF, which is an empirically driven extension to the lowest halo mass rather than the Planck-HMF, although its obvious that the low mass slope is quite shallow and we need more observational constraints which will be possible with the upcoming large spectroscopic surveys such as WAVES \citep{Driver2019}. It is important to note however that the true HMF maybe different from the GSR-HMF, and hence future studies both in theoretical and observational side at lower halo masses are necessary to pin down the exact trend of the HMF. One final reason for preferring the empirical GSR-HMF is that the errors are more realistic whereas the Planck-HMF errors are made artificially small because of the convergence criteria.}


\section{Results with Empirical HMF}
\textcolor{black}{ The baryon density estimates using GSR-HMF and Planck-HMF are shown in Fig.\, \ref{fig:combined_rho_gsr} and \ref{fig:combined_rho_planck} respectively. This involves the same analysis described in Section. 3 but we use the empirical HMF instead of Watson-HMF. The larger uncertainties in the GSR-HMF are reflected in the broader 16th and 84th percentile region in baryon density curves in \ref{fig:combined_rho_gsr} compared to \ref{fig:combined_rho_planck}. }

\textcolor{black}{ Fig.\, \ref{fig:combined_omega_gsr} and \ref{fig:combined_omega_planck} show the cumulative contribution of the baryon density calculated using GSR-HMF and Planck-HMF respectively, by integrating the curves in Fig.\, \ref{fig:combined_rho_gsr} and \ref{fig:combined_rho_planck} from highest to lowest halo mass. The cosmic baryon density estimates using GSR-HMF and Planck-HMF are shown in Table \,\ref{tab:omega}.}

\begin{figure}
 \begin{minipage}{\linewidth}
  \includegraphics[width=\linewidth]{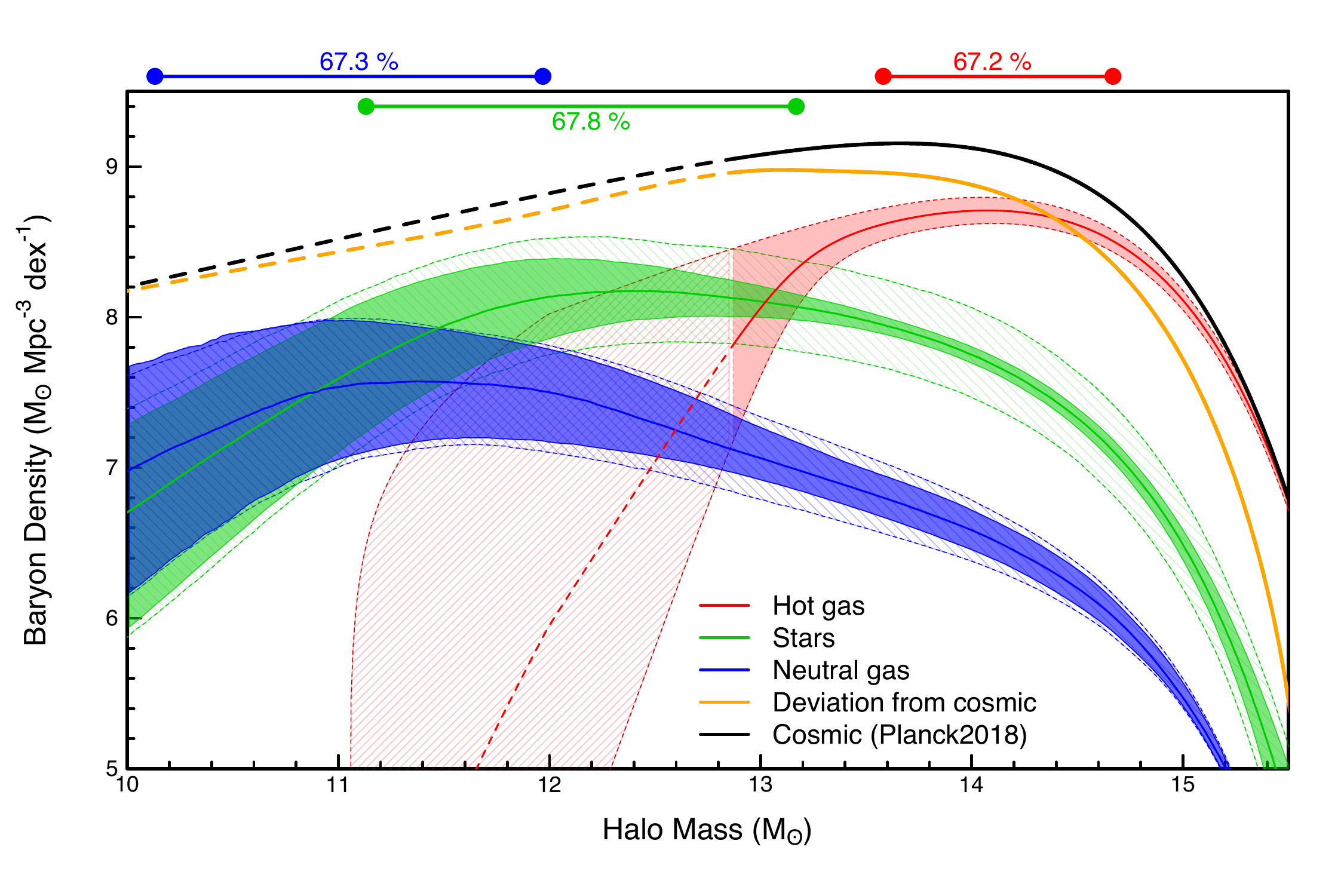}
  \caption{Same as in Fig.\,\ref{fig:combined_rho} but using GSR-HMF. Baryon density for the three components - Neutral gas (in blue), stars (in green) and hot gas (in red). The best-fit scaling relation for the three components have been multiplied by the GSR-HMF to obtain the following baryon density plots.}
  \label{fig:combined_rho_gsr}
 \end{minipage}
 \vspace{1em} 
 \begin{minipage}{\linewidth}
  \includegraphics[width=\linewidth]{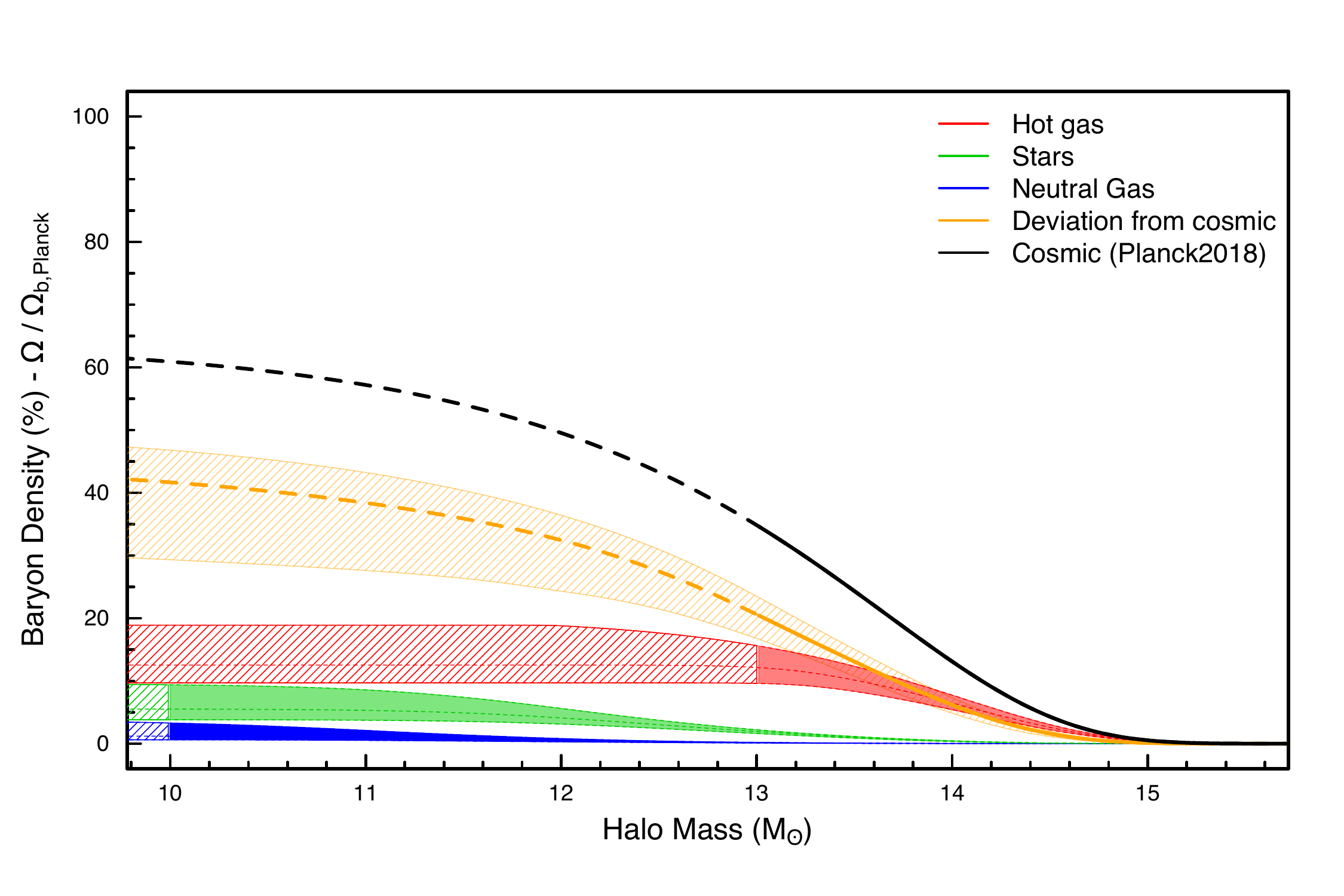}
  \caption{Same as in Fig.\,\ref{fig:combined_omega} but using GSR-HMF. Cumulative contribution of the cosmic baryon density, by integrating from the highest to lowest halo mass, shown as a fraction of the Planck2018 cosmic baryon density value. Cosmic baryon density for the three components - Neutral gas (in blue), stars (in green) and hot gas (in red) are shown. The orange line shows the missing baryon component needed to match the cosmic baryon fraction at each halo mass. The black shows the sum of all the observed and missing baryonic components.}
  \label{fig:combined_omega_gsr}
 \end{minipage}
\end{figure}

\begin{figure}
 \begin{minipage}{\linewidth}
  \includegraphics[width=\linewidth]{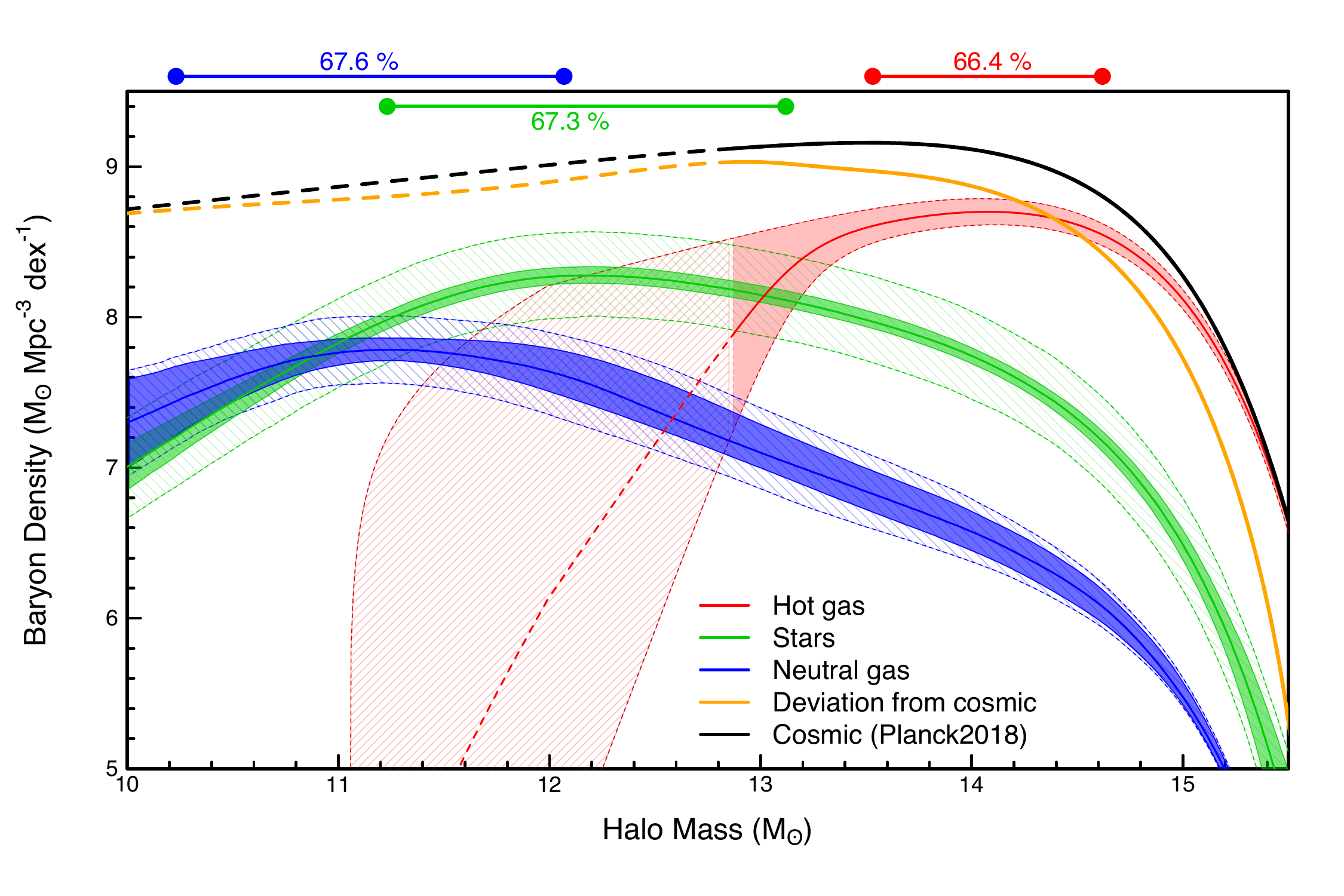}
  \caption{Same as in Fig.\,\ref{fig:combined_rho} but using Planck-HMF. Baryon density for the three components - Neutral gas (in blue), stars (in green) and hot gas (in red). The best-fit scaling relation for the three components have been multiplied by the Planck-HMF to obtain the following baryon density plots.}
  \label{fig:combined_rho_planck}
 \end{minipage}
 \vspace{1em} 
 \begin{minipage}{\linewidth}
  \includegraphics[width=\linewidth]{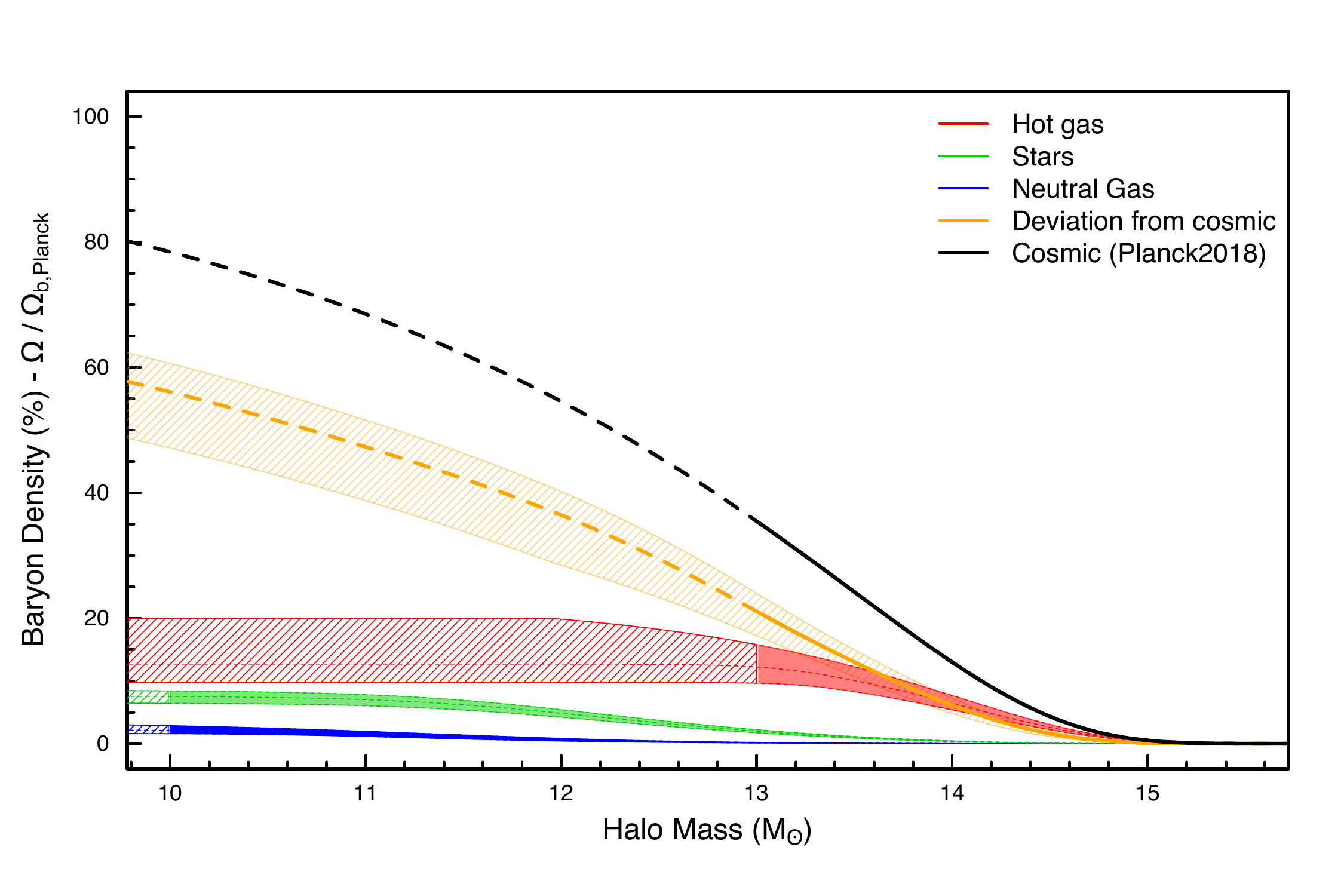}
  \caption{Same as in Fig.\,\ref{fig:combined_omega} but using Planck-HMF. Cumulative contribution of the cosmic baryon density, by integrating from the highest to lowest halo mass, shown as a fraction of the Planck2018 cosmic baryon density value. Cosmic baryon density for the three components - Neutral gas (in blue), stars (in green) and hot gas (in red) are shown. The orange line shows the missing baryon component needed to match the cosmic baryon fraction at each halo mass. The black shows the sum of all the observed and missing baryonic components.}
  \label{fig:combined_omega_planck}
 \end{minipage}
\end{figure}

\bsp	
\label{lastpage}
\end{document}